\documentclass[12pt]{elsarticle}

\usepackage{graphicx}
\usepackage{amsmath}
\usepackage{amsthm}
\usepackage{amsfonts}
\usepackage[utf8]{inputenc}
\usepackage{graphicx}
\usepackage{amsthm}
\usepackage{float}
\usepackage{placeins}

\def\|{\' \i }

\newcounter{bla}

\journal{Computer Physics Communications}

\newtheorem{teor}{Theorem}[section]

\newtheorem{obs}{Remark}[section]
\newtheorem{defin}{Definition}[section]

\begin{document}

\title{\uppercase{Associative Integrator}}


\author[uerj]{J. Avellar}
\ead{jayr.avellar@gmail.com}

\author[uerj]{L.G.S. Duarte}
\ead{lgsduarte@gmail.com.br}

\author[uerj]{A. Fraga}
\ead{alexfisica1976@yahoo.com.br}

\author[uerj]{L.A.C.P. da Mota\corref{cor1}}
\ead{lacpdamota@uerj.br}

\author[uerj]{L. F. de Oliveira}
\ead{lfolive@uerj.br}

\author[uerj,ifrj]{L. O. Pereira}
\ead{leandro.pereira@ifrj.edu.br}

\cortext[cor1]{Corresponding author {\footnotesize
\newline Authors appear in alphabetical order
\newline L.G.S. Duarte and L.A.C.P. da Mota wish to thank Funda\c c\~ao de Amparo \`a Pesquisa do Estado do Rio de Janeiro (FAPERJ) for a Research Grant.}}

\address[uerj]{Universidade do Estado do Rio de Janeiro,
{\it Instituto de F\'{\i}sica, Depto. de F\'{\i}sica Te\'orica},
{\it 20559-900 Rio de Janeiro -- RJ, Brazil}}

\address[ifrj]{ Instituto Federal de Educação Tecnológica do Rio de Janeiro,
{\it  Laboratório de Instrumentação e Simulação Computacional Aplicada LISComp-IFRJ},
{\it Rua Sebastião de Lacerda, Paracambi, RJ, 26600-000, Brazil}}

\begin{abstract}	
Dynamic systems have a fundamental relevance in the description of physical phenomena. The search for more accurate and faster numerical integration methods for the resolution of such systems is, therefore, an important topic of research. The present work introduces a new approach for the numerical integration of dynamic systems. We propose an association of numerical integration methods (integrators)  in order to optimize the performance. The standard we apply is the balance of the duo : {\em precision obtained $\times$ running time}. The numerical integration methods we have chosen, for this particular instance of association, were the Runge-Kutta of fourth order and seventh-eighth order. The algorithm was implemented in $C_{++}$ language. The results showed an improvement in accuracy over the lower grade numerical integrator (actually, we have achieved, basically, the precision of the top integrator) with a processing time  performance closer to the one of the lower grade integrator. Similar results can be obtained for other pairs of numerical integration methods.
\end{abstract}

\begin{keyword}
Numerical Integration, Systems of Ordinary Differential Equations, Runge-Kutta Methods.
\end{keyword}

\maketitle
\section{Introduction}
\label{intro}
\hspace\parindent
In the field of numerical integration, there is a perpetual struggle: running time versus precision of the particular algorithm being used \cite{cellier}. That is a major simplification of the field, of course, but it captures the direction the efforts are made. We want a more precise calculation with the least time to do it.

Surely, for certain applications, the precision is easy to obtain. The data are not so reliable and there is no point in dwelling in having 12 digits of precision. It is not in the nature of the problem to be ``that precise''. But, there is still, even in that situation, the urge to calculate things as fast as one can.

There has been thus produced a plethora \cite{drazin1992nonlinear} of methods and algorithms to solve these kind or riddle: for my particular problem, which method to use? Which algorithm? A Runge-Kutta approach (in which level of precision?)? A Burlisch-Stoer method? And so on. Every problem, every dynamical system, will have their ``best approach`''. We will not enter this debate in detail, try to figure out, analyze types of dynamical systems, etc.

We will introduce a tool that, we hope to demonstrate to the reader, will address that very problem in a general way: It is an way of dealing with a particular dynamical system problem that will provide the researcher with means to fine tuning their numerical integration capabilities to their current interest.

Our research group has been studying Ordinary Differential Equations (ODEs) for a long time \cite{JCAM,AMC,JMP,elementary3D,secondTHEOps1,firsTHEOps1,PS2,nossoPS1CPC} and, in part, analyzing the numerical integration aspects of these equations. Particularly, we have been dedicating sometime to dynamical systems. Either producing computational packages  to calculate the integration of such systems or providing computational tools to analyze different chaotic features such as the fractal dimension of boundaries, the plotting of the solutions in a friendly environment (computer algebra), etc. \cite{ndyn}. We have also dwelt into analyzing the forecast of Time Series. On this latter branch of our research work, we have been concentrating on improving the forecasting capabilities on the global approach to Time Series \cite{imp,new, alt,nc,PRE,carli}.

From this pursuing of better methods and algorithms in the Time Series stage stemmed the method we are hereby introducing. A method designed to work for improving the pure and simple numerical integration, for whatever purpose it is meant to be used.

The paper is divided as follows: In section (\ref{jest}), we introduce the procedure itself apart from elaborate a little more on its motivation and aims. In section (\ref{method}), we describe our proposed new method. We then present (section (\ref{actual})) the results corroborating our claims and analyze the efficiency of the method.

\section{The motivation and the jest of the approach}
\label{jest}

Expanding a little bit on what has already been said on the Introduction of this paper, the main question one tries to address when performing numerical calculations is: what numerical integrator to use to be confident on the results obtained (i.e., the achieved precision is satisfactory) and to obtain these results on the fastest way possible. Which factor contribute to this analysis?

As we have already mentioned, the subject is vast and it is not our intention to cover it all, all its aspects here. We will talk about some general aspects in order to be able to introduce our approach and its limitations.

In general lines, the two main pillars on which one has to base the discussion of precision versus time of computation can be said to be:
\begin{enumerate}
\item {\bf Truncation Error}
\begin{itemize}
\item Basically, its the notion that each particular method of integration is corresponding to a certain order on a Taylor expansion. If one uses up to ``the third order term'', one would have a truncation error of the order of the fourth term, and so on.
\end{itemize}
\item {\bf Round-off Error}
\begin{itemize}
\item  The point here is directly linked to the number of decimal places a computer offers when representing numbers. It is the ``old Grek'' idea that only rational numbers exist. On a computer that is actually true.
\end{itemize}
\end{enumerate}

If the numerical integrator we are using is equivalent to a Taylor series of $n^{th}$ order and with a step $h$, theoretically, the truncation error would be of order $h^{n+1}$. So, of courser, the size of the step $h$ is very relevant on the overall error analysis (even if we are talking about the truncation error only). If we are using a ``high'' value for $n$, $h$ does not need to be small (since $h^{n+1}$ would be small due to the high value for $n$).

So, as we very timidly try to exemplify above, it seems like that all the researcher (that is interested in finding out which numerical procedure to use) has to do is to analyze the precision he/she needs and has (in the original data) and the computer power available in order to decide on which numerical procedure to utilize in order to maximize the ``dynamic duo'': {\bf processing time versus precision} of the procedure. In general, that brings its own perks and difficulties but the point we would like to stress is that the researcher seem to be ``limited'' to the procedures already available and well established. Here, we will introduce an two-fold improvement on that situation: We will produce a procedure that allows one to have higher levels of precision for the same ``time performance'' and also provide a fine-tune tool balancing the ``dynamic duo'' above mentioned.

In a few words, what happened was that, based on our work presented in \cite{PRE}, which was an improvement of the forecast capabilities within the scope of the Global approach to Time Series, we devised a possible new method of proceeding in order to improve the numerical integration that is so useful to us  already.

In \cite{PRE}, we managed to improve the global mapping we were producing by correcting it in order to better predict the ``future'' (unknown) entries of the Time Series using the ``past'' values in the Time Series we had at our disposal. Please see \cite{PRE} for the details.

There are very different characteristics in the two problems: On the Time series scenario, we have a Time Series with (possibly) thousands of entries to fine-tune our Global Mapping but do not know the underlining dynamical system. Whilst, in the numerical integration case, we do know the Dynamical System but want to find its evolution in the future (starting of some initial conditions) from this knowledge. Which are the similarities?

The similar aspect we would like to point out is that, on the Time Series case, we use the Data on the Time Series as the standard to follow, the standard by which we ``correct'' things. On the other hand, on the numerical integration scenario, we use a ``more powerful'' numerical approach in the sense of precision, with the correspondent greater time consumption as the standard by which we will improve a ``lesser'' numerical method (again, in the precision sense), but in a clever way so as not to use so much time as the more precise integrator.

This is the jest of the approach hereby proposed. In the next section, we will introduce it.

\section{The method: The Associative Integration Algorithm - {\bf or AI}}
\label{method}

In this section, we are going to present the algorithm we have developed. This algorithm will prove to be characterized by ensuring the same level of precision of a given method but with fewer steps used in order to obtain that level of precision.

\begin{itemize}
\item First, we will elaborate a method based on the solutions expanded as a Taylor Series. This will set the stage where this Taylor expansion solution will, later on, be replaced by the ``more precise'' algorithm. That will be made clear as we go along (see next item below).
\item Next, we will use the concepts introduced and understood within this context to apply it to a realistic method where the ``perfect'' Taylor expansion approach will be substituted by a practical, well known method. We intend to show that, by using our procedure, we will be able to have a High level of precision with neither any memory problems nor as many steps, leading to an improved algorithm.
\end{itemize}

\subsection{The Taylor Expansion Method}
\label{taylor}

Here, we will briefly introduce the Taylor Expansion approach.

\subsubsection {The Mathematical Basis}

Let us consider the autonomous Dynamical system (in $N$ variable) given by:
\begin{equation}
\label{sistN}
\dot{x}_i = \frac{d x_i}{d t} = f_i(x_1,x_2,\cdots,x_n)\,, \,\,\,\,\,\,(i=1,\cdots,n).
\end{equation}
Consider that $x_i = \phi_i(t)$ are a solution curve (in parametric form) that passes by the point $\vec{x}_0$ for $t=0$. Expanding the functions $\phi_i(t)$ in a Taylor Series, in the point $t=0$, one gets:
\begin{equation}
\label{solN}
x_i = \phi_i(t) = \phi_i(0) + \frac{d\phi_i}{dt}(0)\,t + \frac{d^2\phi_i}{dt^2}(0)\,\frac{t^2}{2!} + \cdots,
\end{equation}
\noindent
Since the Dynamical system is defined by (\ref{sistN}), we have that (over a solution-curve, $x_i = \phi_i(t)$)
 \begin{equation}
\label{solNf}
\dot{x}_i=\dot{\phi}_i(t)=f_i(\vec{x}(t)) \,\,\Rightarrow \,\, \dot{\phi}_i(0)=f_i(\vec{x}(0))=f_i(\vec{x}_0),
\end{equation}
leading that, in second order, one gets:
\begin{equation}
\label{gener}
\frac{d}{dt}\left(\frac{d\phi_i}{dt}\right) = \frac{df_i}{dt} =  \sum_j \frac{\partial f_i}{\partial x_j}\frac{dx_j}{dt} = \sum_j \frac{\partial f_i}{\partial x_j}\,f_j = \sum_j f_j \,\frac{\partial }{\partial x_j}[f_i].
\end{equation}
Since we can write $f_i$ como $\displaystyle{\frac{d x_i}{dt}}$, we have that (suppressing the  $\sum$ for repeated indexes)
\begin{equation}
\label{gener2}
\frac{d}{dt}\left(\frac{d\phi_i}{dt}\right) = f_j \,\frac{\partial }{\partial x_j}\left[f_i\right] =
f_j \,\frac{\partial }{\partial x_j}\left[ \frac{\partial x_i}{\partial x_k}\,\frac{d x_k}{dt}\right] =
f_j \,\frac{\partial }{\partial x_j}\left[ f_k \,\frac{\partial }{\partial x_k}\left[x_i\right]\right].
\end{equation}
So, by defining:
\begin{equation}
\label{generator}
X \equiv f_i \,\frac{\partial }{\partial x_i},
\end{equation}
we can write
\begin{equation}
\label{ddteqX}
\frac{d^u }{dt^u} = X^u.
\end{equation}
Finally, one gets:
\begin{eqnarray}
\label{grupoSeriTaylor}
x_i(t) &=& x_i(0) + t{\it
X}[x_i](0)+{\frac{t^{2}}{2!}} {\it X}^{2}[x_i](0)+ \cdots = \nonumber \\ [3mm]
&=& \left(1+t{\it X}+{\frac{t^{2}}{2!}}{\it X}^{2}+\cdots \right)[x_i](0) = \sum_{k=0}^{\infty}{\frac{t^{k}}{k!}}{\it X}^{k}[x_i](0).
\end{eqnarray}
Using  (\ref{grupoSeriTaylor}) as our basis, we can understand the Taylor Expansion Method for Dynamical Systems. Paying attention to the final result:
\begin{eqnarray}
\label{grupoTaylordt}
x_i(t) = \sum_{k=0}^{\infty}{\frac{t^{k}}{k!}}{\it X}^{k}[x_i](0),
\end{eqnarray}
one can notice that, if one wants to calculate $x_i(\delta t)$, where $\delta t \ll 1$, one can approximate the result by truncating the infinite series in a given order $N$, leading to
\begin{eqnarray}
\label{grupoTaylordtaprox}
x_i(\delta t) \approx \sum_{k=0}^{\scriptstyle{N}}{\frac{{\delta t}^{k}}{k!}}{\it X}^{k}[x_i](0) = F_i(\vec{x}_0;\delta t).
\end{eqnarray}

Theoretically, as big $N$ gets, the better our approximation will be\footnote{For the actual case, for a specific numeric integration of a given dynamical system, , the improvement of the approximation would depend on the number of digits of precision utilized by the computer in question. So, to ensure that with growing $N$ the approximation would improve accordingly, we would have to have a arbitrary digits of precision, as many as needed.}. In this manner, from a given initial point $\vec{x}_0$, we could calculate the following point along the trajectory (which we call  $\vec{x}_1$). From this point,  $\vec{x}_1$ we could (using the same procedure) calculate the next point $\vec{x}_2$, where  $\vec{x}_2 = \vec{F}(\vec{x}_1;\delta t)$, and so on.

In order to make the main idea powering up the method clearer, let us introduce some important results involving solutions to dynamical systems represented by mappings -- please see (\ref{grupoTaylordt}). In what follows, we are going to build a procedure to improve the precision of a given integrator along the lines mention on section (\ref{jest}).

\subsubsection{Main Result}
\label{resultad}

Consider the dynamical system defined by:
\begin{equation}
\label{eqdifdogrupo}
\dot{x_i} = f_i(\vec{x}),\,\,\,\,\, (i=1,\cdots,n),
\end{equation}
with the following solution
\begin{equation}
\label{grupodetransf}
{x_i}(t) = F_i(\vec{x}_0,t),\,\,\,\,\, (i=1,\cdots,n),
\end{equation}
where $f_i(\vec{x}) \equiv {\frac {\partial F_i(\vec{x},t)}{\partial t}}\mid_{t=0}$ and $\dot{x_i} \equiv \frac {d x_i}{d t}$. We saw that the solution  (\ref{grupodetransf}) to the dynamical system (\ref{eqdifdogrupo}) can be obtained from the operator $X \equiv {\sum^n}_{\!\!i=1} \, f_i \,\frac{\partial }{\partial x_i}$:
\begin{equation}
\label{grupoSeriTaylor2}
{x_i}(t) = F_i(\vec{x}_0,t) = x_i(0) + t{\it
X}[x_i](0)+{\frac{t^{2}}{2!}} {\it X}^{2}[x_i](0) + \cdots =
\sum_{k=0}^{\infty}{\frac{t^{k}}{k!}}{\it X}^{k}[x_i](0).
\end{equation}
\noindent
So, starting from a given generic point  $P_0$ (with coordinates $\vec{x}_{(P_0)}$) and working with a time interval  $\delta t$, we can generate a mapping $M$ which takes a point from a given solution-curve to another such point (on the same solution-curve) which correspond to a time increment  $\delta t$ , i.e., to the point that correspond to the solution-curve for the system after the time interval $\delta t$ has passed from the position  $P_0$:
\begin{equation}
\label{mapaSolu}
{x_i}_{(P+\delta P)} = F_i(\vec{x}_{(P)},\delta t) =
\sum_{k=0}^{\infty}{\frac{{\delta t}^{k}}{k!}}{\it X}^{k}[{x_i}]_{(P)}.
\end{equation}
As we have seen, the process of numerically solve the system can be summarized as choosing small time intervals ($\delta t \ll 1$) and truncate the mapping $M$ (\ref{mapaSolu}) in a given order $\scriptstyle{N}$, getting finally the mapping $\overline{M}$ given by:
\begin{equation}
\label{mapaSoluTrunc}
{x_i}_{(P+\overline{\delta P})} = \overline{F}_i(\vec{x}_{(P)},\delta t) =
\sum_{k=0}^{N}{\frac{{\delta t}^{k}}{k!}}{\it X}^{k}[{x_i}]_{(P)},
\end{equation}
\noindent
where ${{x}_i}_{(P+\overline{\delta P})}$ gets close to ${x_i}_{(P+\delta P)}$ when $\delta t \rightarrow 0$. Let us focus on an interesting result:
Consider the functions $\varepsilon_i$ and $\delta^r \varepsilon_i$ defined by
\begin{eqnarray}
\label{varep}
\varepsilon_i(\vec{x})_{(P)} & \equiv & {x_i}_{(P+\delta P)} - {{x}_i}_{(P+\overline{\delta P})} =
\sum_{k=N+1}^{\infty}{\frac{\delta t^{k}}{k!}}{\it X}^{k}[{x_i}]_{(P)} \\[2mm]
\label{dvarep}
\delta \varepsilon_i(\vec{x})_{(P)} & \equiv & \varepsilon_i(\vec{x})_{(P+\delta P)} -
\varepsilon_i(\vec{x})_{(P)} \\[4mm]
\label{dkvarep}
\delta^r \varepsilon_i(\vec{x})_{(P)} & \equiv & \delta^{r-1} \varepsilon_i(\vec{x})_{(P+\delta P)} -
\delta^{r-1} \varepsilon_i(\vec{x})_{(P)}   \,\,\,\,\,\,\, (r=2,\ldots)
\end{eqnarray}
\begin{teor}
\label{teorp}
Consider the dynamical system $\dot{x_i} = f_i(\vec{x})$ with solution  ${x_i}(t) = F_i(\vec{x}_0,t)$ and the mapping $M$ given by  ${x_i}_{(P+\delta P)} = F_i(\vec{x}_{(P)},\delta t) = \sum_{k=0}^{\infty}{\frac{{\delta t}^{k}}{k!}}{\it X}^{k}[{x_i}]_{(P)}$, where the operator $X$ is defined by $X \equiv {\sum^n}_{\!\!i=1} \, f_i \,\frac{\partial }{\partial x_i}$. Also consider that the functions $\varepsilon_i$ and $\delta^r \varepsilon_i$ as defined as above. If the functions  $f_i$ and their derivatives are defined on a point $P$ (with coordinates  $\vec{x}_{(P)}$) and in a non-null neighborhood  arround the point $P$ so we have that:
\begin{equation}
\label{Resultado}
\lim_{\delta t \rightarrow 0} \frac{\delta^{r+1} \varepsilon_i(\vec{x})_{(P)}}{\delta^{r}
\varepsilon_i(\vec{x})_{(P)}} = 0,
\end{equation}
\noindent
where $r$ is  a positive integer
\end{teor}
From equations (\ref{ddteqX}) and (\ref{grupoSeriTaylor}) we can write
\begin{equation}
\label{deltaep}
\delta \varepsilon_i(\vec{x})_{(P)} = \varepsilon_i(\vec{x})_{(P+\delta P)} -
\varepsilon_i(\vec{x})_{(P)} = \sum_{k=N+1}^{\infty}{\frac{\delta t^{k}}{k!}}{\it X}^{k}[{x_i}]_{(P+\delta P)} - \sum_{k=N+1}^{\infty}{\frac{\delta t^{k}}{k!}}{\it X}^{k}[{x_i}]_{(P)}.
\end{equation}
Since ${x_i}_{(P+\delta P)} = {x_i}_{(P)} + \delta {x_i}$, we have (defining ${\Phi^k}_i(\vec{x})_{(P)} = {\it X}^{k}[{x_i}]_{(P)}$):
\begin{eqnarray}
\label{delep}
\delta \varepsilon_i(\vec{x})_{(P)} &=& \sum_{k=N+1}^{\infty}{\frac{\delta t^{k}}{k!}}\left({\it X}^{k}[{x_i}]_{(P+\delta P)} - {\it X}^{k}[{x_i}]_{(P)}\right) =  \nonumber \\ [4mm]
&=& \sum_{k=N+1}^{\infty}{\frac{\delta t^{k}}{k!}}\left({\Phi^k}_i(\vec{x})\mid_{(P+\delta P)}-{\Phi^k}_i(\vec{x})\mid_{(P)}\right) = \nonumber \\ [4mm]
&=& \sum_{k=N+1}^{\infty}{\frac{\delta t^{k}}{k!}}\left( \sum_{j=1}^{n}
\frac{\partial \, {\Phi^k}_i(\vec{x})}{\partial x_j}\mid_{(P)} \, \delta x_j +
\, O({\delta x_j}^{2}) \right),
\end{eqnarray}
\begin{eqnarray}
\label{del2ep}
\delta^2 \varepsilon_i(\vec{x})_{(P)} &=& \sum_{k=N+1}^{\infty}{\frac{\delta t^{k}}{k!}}\left( \sum_{j=1}^{n}
\left(\frac{\partial \, {\Phi^k}_i(\vec{x})}{\partial x_j}\mid_{(P+\delta P)} -
\frac{\partial \, {\Phi^k}_i(\vec{x})}{\partial x_j}\mid_{(P)} \right)\, \delta x_j +
\, O({\delta x_j}^{2}) \right) =  \nonumber \\ [4mm]
&=& \sum_{k=N+1}^{\infty}{\frac{\delta t^{k}}{k!}}\left( \sum_{j=1}^{n}
\frac{\partial^2 \, {\Phi^k}_i(\vec{x})}{\partial x_l \partial x_j}\mid_{(P)}\, \delta x_l \, \delta x_j+
\, O({\delta x}^{3}) \right),
\end{eqnarray}
$\cdots$
\begin{equation}
\label{delrep}
\delta^r \varepsilon_i(\vec{x})_{(P)} = \sum_{k=N+1}^{\infty}{\frac{\delta t^{k}}{k!}}\left( \sum_{j_1,\cdots,j_r=1}^{n}
\frac{\partial^r \, {\Phi^k}_i(\vec{x})}{\partial x_{j_1} \cdots \partial x_{j_r}}\mid_{(P)}\, \delta x_{j_l} \cdots \delta x_{j_r}+
\, O({\delta x}^{r+1}) \right).
\end{equation}
So,
\begin{equation}
\label{delrm1sdelr}
\lim_{\delta t \rightarrow 0} \frac{\delta^{r+1} \varepsilon_i}{\delta^{r} \varepsilon_i} =
\frac{\sum_{k=N+1}^{\infty}{\frac{\delta t^{k}}{k!}}\left( \sum_{j_1,\cdots,j_{r+1}=1}^{n}
\frac{\partial^{r+1} \, {\Phi^k}_i(\vec{x})}{\partial x_{j_1} \cdots \partial x_{j_{r+1}}}\mid_{(P)}\, \delta x_{j_l} \cdots \delta x_{j_{r+1}}+\, O({\delta x}^{r+2}) \right)}
{\sum_{k=N+1}^{\infty}{\frac{\delta t^{k}}{k!}}\left( \sum_{j_1,\cdots,j_r=1}^{n}
\frac{\partial^r \, {\Phi^k}_i(\vec{x})}{\partial x_{j_1} \cdots \partial x_{j_r}}\mid_{(P)}\, \delta x_{j_l} \cdots \delta x_{j_r}+\, O({\delta x}^{r+1}) \right)}.
\end{equation}

Since we are on regular points of the system, we have  $\lim_{\delta t \rightarrow 0}\, \frac{\delta x}{\delta t}=K$, where $K$ is an $n-$upla of finite real numbers and, therefore, $\delta t \rightarrow 0$ implies that  $\delta x \rightarrow K\,\delta t$ which, in turn, implies that the above limit is zero.

In the next subsection, we will show that, based on this result above, there is a way to associate two integrators,  each with a different precision capability, in such a manner as to have the processing time closer to the range of the lesser (in the sense of precision) method and the precision closer to the more efficient method (in the sense of precision yet again).

\subsection{The structural basis of our AI approach}
\label{AlgIntAss}

Let us consider the low dimension dynamical system, i.e., $\dot{\vec{x}}=\vec{f}(\vec{x})$ where the dimension of the vector  $\vec{x}$ is an integer  $n$ no too big\footnote{The term ``not too big'' deserves a better explanation: In practice, it means not bigger than 12}. We have seen above that we can expand the solutions as a Taylor Series and, for a given time interval $\delta t$, buil a mapping given by:
$$
{x_i}_{(P+\delta P)} = \sum_{k=0}^{\infty}{\frac{{\delta t}^{k}}{k!}}{\it X}^{k}[x_i]_{(P)} = F_i(\vec{x}_{(P)};\delta t),
$$
which, starting from a given point $P_0$ (with coordinates $\vec{x}_0$), calculates the following points of a solution-curve (passing through $P_0$) separated by a time interval $\delta t$. In theory, the mapping  $M$ would do that with infinite precision. Since it is obviously not possible due to the fact that it would imply the use of an infinite number of terms,  what we do in practice is truncate the series at some finite order $\scriptstyle{N}$. Suppose we have done that and that we have chosen  $\delta t \ll 1$ (disregarding higher order terms). We would have obtained thus a mapping  $\overline{M}$, with precision corresponding to a certain number of digits $D$. Consider that we chose an initial point  $P_0$ and that we have access to a certain number of points in the solution-curve obtained with an arbitrary precision $Pr_A \gg D$. Consider also that the function $\delta^r \varepsilon_i$ (uma função $\varepsilon_i$ corresponds to $r=0$) on the points following the point $P_0$, produced by the mapping  ${M}$. One can infer that:

\begin{obs}
Since  $\delta t \ll 1$ and the mapping  $\overline{M}$ corresponds to the truncation of the mapping  $M$ in the order  $\scriptstyle{N}$, there exists a maximum  integer positive  number  $\scriptstyle{L}$ such that
\begin{equation}
\label{Resulpratico}
\frac{\delta^{L} \varepsilon}{\delta^{L-1} \varepsilon} < 1.
\end{equation}
\end{obs}
\begin{obs}
The number $\scriptstyle{L}$ grows as the number  $\scriptstyle{N}$ grows (the order of truncation) and as the number $\delta t$ diminishes (the time interval between two sequential points generated by the mapping ${M}$).
\end{obs}
\begin{obs}
As a consequence of the fact that $\delta t \ll 1$ and from theorem \ref{teorp}, the moduli of the functions  $\delta^r \varepsilon_i$ are $\ll 1$.
\end{obs}

\bigskip

Let us now examine the functions $\delta^r \varepsilon_i$ on the point $P_0$ and the following ones, generated by the mapping  $M$. In $P_0$, we have:
\begin{equation}
\label{d0ei0}
\delta^0 \varepsilon_i(\vec{x})_{(P_0)} = \varepsilon_i(\vec{x})_{(P_0)} = {x_i}_{(P_0+\delta P_0)} - {x_i}_{(P_0+\overline{\delta P}_0)} = \sum_{k=N+1}^{\infty}{\frac{\delta t^{k}}{k!}}{\it X}^{k}[{x_i}]_{(P_0)}.
\end{equation}
Simplifying the notation, let us call $P_1 \equiv P_0+{\delta P}_0$ and $\overline{P}_1 \equiv P_0+\overline{\delta P}_0$, and, analogously, $P_{j+1} \equiv P_j+{\delta P}_j$ e $\overline{P}_{j+1} \equiv P_j+\overline{\delta P}_j$. Consider now the  $p$ points after $P_0$:
\begin{equation}
\label{d0ei1}
\delta^0 \varepsilon_i(\vec{x})_{(P_1)} = \varepsilon_i(\vec{x})_{(P_1)} = {x_i}_{(P_1+\delta P_1)} - {x_i}_{(P_1+\overline{\delta P}_1)} = \sum_{k=N+1}^{\infty}{\frac{\delta t^{k}}{k!}}{\it X}^{k}[{x_i}]_{(P_1)}.
\end{equation}
$\cdots$
\begin{equation}
\label{d0eipm1}
\delta^0 \varepsilon_i(\vec{x})_{(P_{p-1})} = \varepsilon_i(\vec{x})_{(P_{p-1})} = {x_i}_{(P_{p-1}+\delta P_{p-1})} - {x_i}_{(P_{p-1}+\overline{\delta P}_{p-1})} = \sum_{k=N+1}^{\infty}{\frac{\delta t^{k}}{k!}}{\it X}^{k}[{x_i}]_{(P_{p-1})}.
\end{equation}
For the functions  $\delta \varepsilon_i$, we have:
\begin{equation}
\label{d1ei0}
\delta \varepsilon_i(\vec{x})_{(P_0)} = \varepsilon_i(\vec{x})_{(P_0+\delta P_0)} - \varepsilon_i(\vec{x})_{(P_0)},
\end{equation}
\begin{equation}
\label{d1ei1}
\delta \varepsilon_i(\vec{x})_{(P_1)} = \varepsilon_i(\vec{x})_{(P_1+\delta P_1)} - \varepsilon_i(\vec{x})_{(P_1)},
\end{equation}
$\cdots$
\begin{equation}
\label{d1eipm1}
\delta \varepsilon_i(\vec{x})_{(P_{p-1})} = \varepsilon_i(\vec{x})_{(P_{p-1}+\delta P_{p-1})} - \varepsilon_i(\vec{x})_{(P_{p-1})},
\end{equation}
One can write for the functions $\delta^r \varepsilon_i$,
\begin{equation}
\label{dreipm1}
\delta^r \varepsilon_i(\vec{x})_{(P_{j-1})} = \delta^{r-1} \varepsilon_i(\vec{x})_{(P_{j-1}+\delta P_{j-1})} - \delta^{r-1} \varepsilon_i(\vec{x})_{(P_{j-1})}. (r=2)
\end{equation}
In a more detailed analysis, we observe a regular behaviour of the function  $\delta^r \varepsilon_i(\vec{x})_{(P_i)}$. In order to do that, let us consider that $P=4 r=3$, so one gets:

\bigskip
$\delta^0\varepsilon_i(\vec{x})_{(4)} =
\varepsilon_i(\vec{x})_{(4)}$

$\delta^1\varepsilon_i(\vec{x})_{(4)} = \delta^0\varepsilon_i(\vec{x})_{(4)}-\delta^0\varepsilon_i(\vec{x})_{(3)}$

$\delta^2\varepsilon_i(\vec{x})_{(4)} =\delta^1\varepsilon_i(\vec{x})_{(4)}-\delta^1\varepsilon_i(\vec{x})_{(3)}=(\varepsilon_i(\vec{x})_{(4)}-\varepsilon_i(\vec{x})_{(3)})-(\varepsilon_i(\vec{x})_{(3)}-\varepsilon_i(\vec{x})_{(2)})= \varepsilon_i(\vec{x})_{(4)}-2\varepsilon_i(\vec{x})_{(3)}+\varepsilon_i(\vec{x})_{(2)}$

$\delta^3\varepsilon_i(\vec{x})_{(4)} =\delta^2\varepsilon_i(\vec{x})_{(4)}-\delta^2\varepsilon_i(\vec{x})_{(3)}=(\delta^1\varepsilon_i(\vec{x})_{(4)}-\delta^1\varepsilon_i(\vec{x})_{(3)})-(\delta^1\varepsilon_i(\vec{x})_{(3)}-\delta^1\varepsilon_i(\vec{x})_{(2)})=(\varepsilon_i(\vec{x})_{(4)}-2\varepsilon_i(\vec{x})_{(3)})+(\varepsilon_i(\vec{x})_{(2)})-[(\varepsilon_i(\vec{x})_{(3)}-\varepsilon_i(\vec{x})_{(2)})-(\varepsilon_i(\vec{x})_{(2)}-\varepsilon_i(\vec{x})_{(1)})]= \varepsilon_i(\vec{x})_{(4)}-3\varepsilon_i(\vec{x})_{(3)}+3\varepsilon_i(\vec{x})_{(2)}-\varepsilon_i(\vec{x})_{(1)}$

\bigskip
Organizing the result, one has:\\

$\delta^0\varepsilon_i(\vec{x})_{(4)} = \textbf{1}\varepsilon_i(\vec{x})_{(4)}$

$\delta^1\varepsilon_i(\vec{x})_{(4)} = \textbf{1}\varepsilon_i(\vec{x})_{(4)}-\textbf{1}\varepsilon_i(\vec{x})_{(3)}$

$\delta^2\varepsilon_i(\vec{x})_{(4)} = \textbf{1}\varepsilon_i(\vec{x})_{(4)}-\textbf{2}\varepsilon_i(\vec{x})_{(3)}+\textbf{1}\varepsilon_i(\vec{x})_{(2)}$

$\delta^3\varepsilon_i(\vec{x})_{(4)} = \textbf{1}\varepsilon_i(\vec{x})_{(4)}-\textbf{3}\varepsilon_i(\vec{x})_{(3)}+\textbf{3}\varepsilon_i(\vec{x})_{(2)}-\textbf{1}\varepsilon_i(\vec{x})_{(1)}$.\\

We can them generalize the function $\delta^r \varepsilon_i(\vec{x})_{(P_i)}$
such that, for each value of  $r$, we have as coefficients the terms of a line of the triangle of Pascal for the values of $\varepsilon_i(\vec{x})_{(p-j)}$.

\bigskip
\bigskip

$\delta^0\varepsilon_i(\vec{x})_{(p)} = \textbf{1}\varepsilon_i(\vec{x})_{(p)}$

$\delta^1\varepsilon_i(\vec{x})_{(p)} = \textbf{1}\varepsilon_i(\vec{x})_{(p)}-\textbf{1}\varepsilon_i(\vec{x})_{(p-1)}$

$\delta^2\varepsilon_i(\vec{x})_{(p)} = \textbf{1}\varepsilon_i(\vec{x})_{(p)}-\textbf{2}\varepsilon_i(\vec{x})_{(p-1)}+\textbf{1}\varepsilon_i(\vec{x})_{(p-2)}$

$\delta^3\varepsilon_i(\vec{x})_{(p)} = \textbf{1}\varepsilon_i(\vec{x})_{(p)}-\textbf{3}\varepsilon_i(\vec{x})_{(p-1)}+\textbf{3}\varepsilon_i(\vec{x})_{(p-2)}-\textbf{1}\varepsilon_i(\vec{x})_{(p-3)}$

$\delta^4\varepsilon_i(\vec{x})_{(p)} = \textbf{1}\varepsilon_i(\vec{x})_{(p)}-\textbf{4}\varepsilon_i(\vec{x})_{(p-1)}+\textbf{6}\varepsilon_i(\vec{x})_{(p-2)}-\textbf{4}\varepsilon_i(\vec{x})_{(p-3)}+\textbf{1}\varepsilon_i(\vec{x})_{(p-4)}$\\

The general term is given by:
\begin{equation}
\label{eq42}
\delta^r \varepsilon_i(\vec{x})_{(p)} = \sum_{j=0}^{r}{\left(
\begin{array}{c}
r\\
j
\end{array}
\right)(-1)^{(r-j)}\varepsilon_i(\vec{x})_{(p-j)}
},
\end{equation}
\medskip

Now, we have the theoretical basis in which  to build the central ideia for our new algorithm. Consider that we have  $p$ points $P_1, P_2, \ldots, P_p$, generated from  $P_0$ by the mapping $M$ (with and arbitrary precision that, for the moment, we will regard as unbounded). One can observe that:

\begin{obs}
\label{dtm1}
 Let $\delta t \ll 1$, so the point  $P_1$ is very close to the point $P_0$ \footnote{We are considering the usual metric, i.e., the distance from  $P_0$ to $P_1$ is given by $dd(P_0,P_1) \equiv  \displaystyle{\sqrt{\sum (\delta {x_i}_{(P_0)})^2}}$.}, in other words, the distance from  $P_0$ to $P_1$ is much less than $1$ ($dd(P_0,P_1) \ll 1$). That implies that  $|\delta {x_i}| \ll 1$. Furthermore, we can infer:
\begin{eqnarray}
{x_i}_{(P+\delta P)} &=& \sum_{k=0}^{\infty}{\frac{{\delta t}^{k}}{k!}}{\it X}^{k}[x_i]_{(P)} = {x_i}_{(P)} + \sum_{k=1}^{\infty}{\frac{{\delta t}^{k}}{k!}}{\Phi^k}_i(\vec{x})_{(P)}, \nonumber \\
& \Rightarrow & {x_i}_{(P+\delta P)} - {x_i}_{(P)} = \delta {x_i} = {\Phi^1}_i(\vec{x})_{(P)} \delta t + \sum_{k=2}^{\infty}{\frac{{\delta t}^{k}}{k!}}{\Phi^k}_i(\vec{x})_{(P)} \nonumber \\
& \Rightarrow & O({\delta {x_i}}) \approx O({\delta t}).
\end{eqnarray}

\end{obs}

\begin{obs}
\label{ordin}
From the definition of the functions  $\delta^r \varepsilon_i$, we get:
\begin{itemize}
\item From (\ref{d0ei0}) we get $\varepsilon_i(\vec{x})_{(P)} = \sum_{k=N+1}^{\infty}{\frac{\delta t^{k}}{k!}}{{\Phi^k}_i}(\vec{x})_{(P)} \approx O({\delta t}^{N+1})$.
\item From (\ref{deltaep},\ref{d1ei0}) we can conclude
$$
\delta \varepsilon_i(\vec{x})_{(P)} = \varepsilon_i(\vec{x})_{(P+\delta P)} - \varepsilon_i(\vec{x})_{(P)} = \sum_{k=N+1}^{\infty}{\frac{\delta t^{k}}{k!}}\left( \sum_{j=1}^{n}
\frac{\partial \, {\Phi^k}_i(\vec{x})}{\partial x_j}\mid_{(P)} \, \delta x_j +
\, O({\delta x_j}^{2}) \right),
$$
$\Rightarrow\, \delta \varepsilon_i(\vec{x})_{(P)} \approx O({\delta t}^{N+2})$.
\item Finally, from (\ref{delrep}) we get
$$
\delta^r \varepsilon_i(\vec{x})_{(P)} = \sum_{k=N+1}^{\infty}{\frac{\delta t^{k}}{k!}}\left( \sum_{j_1,\cdots,j_r=1}^{n} \frac{\partial^r \, {\Phi^k}_i(\vec{x})}{\partial x_{j_1} \cdots \partial x_{j_r}}\mid_{(P)}\, \delta x_{j_l} \cdots \delta x_{j_r}+ \, O({\delta x}^{r+1}) \right),
$$
$\Rightarrow\, \delta^r \varepsilon_i(\vec{x})_{(P)} \approx O({\delta t}^{N+r+1})$.
\end{itemize}
In words, for each  $\delta^r \varepsilon_i(\vec{x})_{(p)}$ we use, we get an improvement on the precision of order  $O({\delta t})$.
\end{obs}

Let us suppose that, from a given point $P_0$, we apply the mapping  $M$ to calculate the next $p$ points $P_1,\ldots,P_p$ and, after those we use the mapping $\overline{M}$ to calculate  $p+1$ points $\overline{P}_1,\ldots,\overline{P}_{p+1}$. Let us now suppose we want to calculate the point ${P}_{p+1}$, {\bf without} using the mapping $M$, i.e., we want to approximate, we want to calculate  ${x_i}_{(P_{p+1})}$. We know from  (\ref{d0eipm1}) that $\varepsilon_i(\vec{x})_{(P_{p})} = {x_i}_{(P_{p+1})} - {x_i}_{(\overline{P}_{p+1})}$, and that implies that  ${x_i}_{(P_{p+1})} = {x_i}_{(\overline{P}_{p+1})} + \varepsilon_i(\vec{x})_{(P_{p})}$. Therefore, if we knew  $\varepsilon_i(\vec{x})_{(P_{p})}$, we would be able to calculate ${x_i}_{(P_{p+1})}$. We also know that, from (\ref{d1eipm1}),  $\delta \varepsilon_i(\vec{x})_{(P_{p-1})} = \varepsilon_i(\vec{x})_{(P_{p})} - \varepsilon_i(\vec{x})_{(P_{p-1})}$ and, so, $\varepsilon_i(\vec{x})_{(P_{p})} = \varepsilon_i(\vec{x})_{(P_{p-1})} + \delta \varepsilon_i(\vec{x})_{(P_{p-1})}$. This finally leads to
\begin{equation}
\label{primapro}
{x_i}_{(P_{p+1})} = {x_i}_{(\overline{P}_{p+1})} + \varepsilon_i(\vec{x})_{(P_{p-1})} + \delta \varepsilon_i(\vec{x})_{(P_{p-1})}.
\end{equation}
Please note that the two first terms on the right-hand side of (\ref{primapro}) are known and, besides that, their sum (${x_i}_{(\overline{P}_{p+1})} + \varepsilon_i(\vec{x})_{(P_{p-1})}$) represents a better approximation than  ${x_i}_{(\overline{P}_{p+1})}$ to the value of ${x_i}_{(P_{p+1})}$. In order to better understand that, one can look on the analysis we presented on remarks \ref{dtm1} and \ref{ordin}:
\begin{equation}
\label{oepdep}
\varepsilon_i(\vec{x})_{(P)}  \approx O({\delta t}^{N+1}), \,\,
\delta \varepsilon_i(\vec{x})_{(P)} \approx O({\delta t}^{N+2}).
\end{equation}
So, the error on the approximation when we use  ${x_i}_{(\overline{P}_{p+1})}$ has order $O({\delta t}^{N+1})$, on the other hand, the approximation error when using  ${x_i}_{(\overline{P}_{p+1})} + \varepsilon_i(\vec{x})_{(P_{p-1})}$ is of order $O({\delta t}^{N+2})$. This reasoning can be extended: we do not know the value of $\delta\varepsilon_i(\vec{x})_{(P_{p-1})}$, but we know that $\delta^2 \varepsilon_i(\vec{x})_{(P_{p-2})} = \delta \varepsilon_i(\vec{x})_{(P_{p-1})} - \delta \varepsilon_i(\vec{x})_{(P_{p-2})}$ and, therefore, $\delta \varepsilon_i(\vec{x})_{(P_{p-1})} = \delta \varepsilon_i(\vec{x})_{(P_{p-2})} + \delta^2 \varepsilon_i(\vec{x})_{(P_{p-2})}$, which implies that
\begin{equation}
\label{segapro}
{x_i}_{(P_{p+1})} = {x_i}_{(\overline{P}_{p+1})} + \varepsilon_i(\vec{x})_{(P_{p-1})} + \delta \varepsilon_i(\vec{x})_{(P_{p-2})} + \delta^2 \varepsilon_i(\vec{x})_{(P_{p-2})}.
\end{equation}
Analogously, we do not know  $\delta^2 \varepsilon_i(\vec{x})_{(P_{p-2})}$ but, since $\delta^2 \varepsilon_i(\vec{x})_{(P)} \approx O({\delta t}^{N+3})$, approximating  ${x_i}_{(P_{p+1})}$ by
\begin{equation}
\label{segapro2}
{x_i}_{(P_{p+1})} \approx {x_i}_{(\overline{P}_{p+1})} + \varepsilon_i(\vec{x})_{(P_{p-1})} + \delta \varepsilon_i(\vec{x})_{(P_{p-2})},
\end{equation}
is better than using
\begin{equation}
\label{primapro2}
{x_i}_{(P_{p+1})} \approx {x_i}_{(\overline{P}_{p+1})} + \varepsilon_i(\vec{x})_{(P_{p-1})}.
\end{equation}

We can go on with this line of reasoning and that leads to:
\begin{equation}
\label{pesiap}
{x_i}_{(P_{p+1})} = {x_i}_{(\overline{P}_{p+1})} + \varepsilon_i(\vec{x})_{(P_{p-1})} + \delta \varepsilon_i(\vec{x})_{(P_{p-2})} + \delta^2 \varepsilon_i(\vec{x})_{(P_{p-3})} + \cdots + \delta^r \varepsilon_i(\vec{x})_{(P_{p-r-1})} + \cdots.
\end{equation}
Using the results from equation (\ref{eq42}) $
\delta^r \varepsilon_i(\vec{x})_{(p)} = \sum_{j=0}^{r}{\left(
\begin{array}{c}
r\\
j
\end{array}
\right)(-1)^{(r-j)}\varepsilon_i(\vec{x})_{(p-j)}
}$, we can re-write equation  \ref{pesiap}  as:
\begin{equation}
\label{pesiapc}
{x_i}_{(P_{p+1})} = {x_i}_{(\overline{P}_{p+1})}+\sum_{k=0}^{r}\delta^k\varepsilon_i(\vec{x})_{(p)}.
\end{equation}
In the next subsection, we will demonstrate how to use the results  (\ref{pesiapc}) to build an algorithm for numerical integration.

\subsection{The Algorithm itself}
\label{passosAlg}

The mapping  $M$ defined on the previous subsection can not be used in practice since it uses infinite operations. But, surely, since $\scriptstyle{N}$ is a finite integer, one can use mapping $\overline{M}$, defined by ${x_i}_{(P+\overline{\delta P})} = \sum_{k=0}^{N}{\frac{{\delta t}^{k}}{k!}}{\it X}^{k}[{x_i}]_{(P)}$,  in a practical way. Let us now see how we can use the above idea of having two mappings (of two different precision levels) in an associative relation to build our algorithm.

\begin{defin}
\label{Mdef}
Consider the two mappings $M^{+}$ e $M^{-}$ defined by:
\begin{eqnarray}
M^{+} &\equiv& {x_i}_{(P+\overline{\delta P})} = \sum_{k=0}^{N^{+}}{\frac{{\delta t}^{k}}{k!}}{\it X}^{k}[{x_i}]_{(P)}, \\
M^{-} &\equiv& {x_i}_{(P+\overline{\delta P})} = \sum_{k=0}^{N^{-}}{\frac{{\delta t}^{k}}{k!}}{\it X}^{k}[{x_i}]_{(P)},
\end{eqnarray}
where $N^{+}$ and $N^{-}$ are two positive integers that obey $N^{+} > N^{-}$.
\end{defin}

The basic idea is to use the more precise mapping ($M^{+}$) to play the role of the mapping  $M$ (as above) and the less precise mapping ($M^{-}$) in the role of the mapping $\overline{M}$. Please note that, in this ``real'' situation we are presenting now, both the sets of  de $n-$uplas that represent the points  $P^{+}$ and $P^{-}$ present limit digits (precision) and, therefore, the order of the approximation, that will be the analogous of the approximation (\ref{pesiapc}), would be of a finite order: $r$.  Using this idea together with the results presented in the last subsection, one can build an algorithm that will use the strength of both mappings $M^{+}$ e $M^{-}$. In order to achieve that, let us define the functions $\Delta^k \varepsilon_i$ (the analogue to the functions (\ref{dkvarep})), as follows:

\begin{defin}
\label{Deledef}
Let us consider a point  $P_0$ and, using the mapping  $M^{+}$ as defined above, calculate a certain number $p$ of points that follow this point  $P_0$. Let us call them  $P^{+}_1,\,P^{+}_2,\,\ldots,\,P^{+}_p$ (where $P^{+}_{1} \equiv M^{+}[P_0]$ and $P^{+}_{j+1} \equiv M^{+}[P^{+}_j]$). Using the mapping $M^{-}$ as defined above, we can, using the points  $P_0,\,P^{+}_1,\,P^{+}_2,\,\ldots,\,P^{+}_p$, determine $p+1$ points defined by:
\begin{equation}
\label{pmenos}
P^{-}_1 \equiv M^{-}[P_0],\,\,\,P^{-}_2 \equiv M^{-}[P^{+}_1],\,\,\,P^{-}_3 \equiv M^{-}[P^{+}_2],\ldots, P^{-}_{p+1} \equiv M^{-}[P^{+}_p].
\end{equation}
From the points $P_0$, $P^{+}$ e $P^{-}$, one can define functions $\Delta^r \varepsilon_i$ (the analogue to the functions (\ref{dkvarep})) as:
\begin{eqnarray}
\label{varepReal}
\Delta^0 \varepsilon_i(\vec{x})_{(P^{+}_j)} & \equiv & {x_i}_{(P^{+}_j)} - {x_i}_{(P^{-}_j)}, \,\, (j=1,\ldots,p), \nonumber \\ [2mm]
\Delta^1 \varepsilon_i(\vec{x})_{(P^{+}_j)} & \equiv & \Delta^0 \varepsilon_i(\vec{x})_{(P^{+}_j)} -
\Delta^0 \varepsilon_i(\vec{x})_{(P^{+}_{j-1})}, \,\, (j=2,\ldots,p), \nonumber \\
\vdots & & \vdots \nonumber \\
\Delta^r \varepsilon_i(\vec{x})_{(P^{+}_j)} & \equiv & \Delta^{r-1} \varepsilon_i(\vec{x})_{(P^{+}_j)} -
\Delta^{r-1} \varepsilon_i(\vec{x})_{(P^{+}_{j-1})}, \,\, (j=r+1,\ldots,p), \label{Drepi} \\
\end{eqnarray}
\end{defin}
then, analogously  to  equation (\ref{eq42}) leads to:
\begin{equation}
\label{eq56}
\Delta^r \varepsilon_i(\vec{x})_{(P^+_p)} = \sum_{j=0}^{r}{\left(
\begin{array}{c}
r\\
j
\end{array}
\right)(-1)^{(r-j)}\varepsilon_i(\vec{x})_{(P^+_{p-j})}
},
\end{equation}

\noindent
From the functions $\Delta^r \varepsilon_i$ as defined above and from the points  $P_0$, $P^{+}$ e $P^{-}$ we can calculate an approximation to the  point $P^{+}_{p+1}$ without using the mapping  $M^{+}$. Let us star with an approximation of order ``zero'':
\begin{equation}
\label{apro0}
{x_i}_{(P^{+}_{p+1})} \approx {x_i}_{(P^{-}_{p+1})},
\end{equation}
where the order of the approximations is, according to what we show on the remark \ref{ordin}, $O({\delta t}^{(N^{+}-N^{-}+1)})$.
But we know that ${x_i}_{(P^{+}_{p+1})} - {x_i}_{(P^{-}_{p+1})} = \Delta^0
\varepsilon_i(\vec{x})_{(P^{+}_{p+1})}$ e $\Delta^1 \varepsilon_i(\vec{x})_{(P^{+}_{p+1})} = \Delta^0 \varepsilon_i(\vec{x})_{(P^{+}_{p+1})} - \Delta^0 \varepsilon_i(\vec{x})_{(P^{+}_p)}$, therefore,
\begin{equation}
\label{equo1}
{x_i}_{(P^{+}_{p+1})} = {x_i}_{(P^{-}_{p+1})} + \Delta^0 \varepsilon_i(\vec{x})_{(P^{+}_p)} + \Delta^1 \varepsilon_i(\vec{x})_{(P^{+}_{p+1})},
\end{equation}
thus leading to the approximation of ``order one'':
\begin{equation}
\label{apro1}
{x_i}_{(P^{+}_{p+1})} \approx {x_i}_{(P^{-}_{p+1})} + \Delta^0 \varepsilon_i(\vec{x})_{(P^{+}_p)},
\end{equation}
where the order of the approximation is $O({\delta t}^{(N^{+}-N^{-}+2)})$. But we know that $\Delta^2 \varepsilon_i(\vec{x})_{(P^{+}_{p+1})} =$ $\Delta^1 \varepsilon_i(\vec{x})_{(P^{+}_{p+1})} - \Delta^1 \varepsilon_i(\vec{x})_{(P^{+}_p)}$ and, so,
\begin{equation}
\label{equo2}
{x_i}_{(P^{+}_{p+1})} = {x_i}_{(P^{-}_{p+1})} + \Delta^0 \varepsilon_i(\vec{x})_{(P^{+}_p)} + \Delta^1 \varepsilon_i(\vec{x})_{(P^{+}_{p})} + \Delta^2 \varepsilon_i(\vec{x})_{(P^{+}_{p+1})},
\end{equation}
leading to the ``order 2'' for the approximation:
\begin{equation}
\label{apro2}
{x_i}_{(P^{+}_{p+1})} \approx {x_i}_{(P^{-}_{p+1})} + \Delta^0 \varepsilon_i(\vec{x})_{(P^{+}_p)} + \Delta^1 \varepsilon_i(\vec{x})_{(P^{+}_{p})},
\end{equation}
where the order of the approximation is $O({\delta t}^{(N^{+}-N^{-}+3)})$. Surely, we can go on with this reasoning and, ingeneral, we would have:
\begin{equation}
\label{equor}
{x_i}_{(P^{+}_{p+1})} = {x_i}_{(P^{-}_{p+1})} + \Delta^0 \varepsilon_i(\vec{x})_{(P^{+}_p)} + \cdots + \Delta^{r-1} \varepsilon_i(\vec{x})_{(P^{+}_{p})} + \Delta^r \varepsilon_i(\vec{x})_{(P^{+}_{p+1})},
\end{equation}
Leading to the approximation of  ``order $r$'':
\begin{equation}
\label{apror}
{x_i}_{(P^{+}_{p+1})} \approx {x_i}_{(P^{-}_{p+1})} + \Delta^0 \varepsilon_i(\vec{x})_{(P^{+}_p)} + \cdots + \Delta^{r-1} \varepsilon_i(\vec{x})_{(P^{+}_{p})},
\end{equation}
that, according to equation  (\ref{eq56}) leads to:
\begin{equation}
\label{eq64}
{x_i}_{(P^+_{p+1})} = {x_i}_{(P^-_{p+1})}+\sum_{k=0}^{r}\Delta^k\varepsilon_i(\vec{x})_{(P^+_p)}.
\end{equation}
where the approximation order is $O({\delta t}^{(N^{+}-N^{-}+r+1)})$ and $r \leq p$.\\

The figure (\ref{fig:fig1}) shows a diagram for the algorithm, from now on called {\bf ``the stitching process''}:

\bigskip\bigskip\bigskip\bigskip\bigskip\bigskip\bigskip\bigskip\bigskip\bigskip
\begin{figure}[H]
\includegraphics[scale=0.45,bb=0 0 3 3]{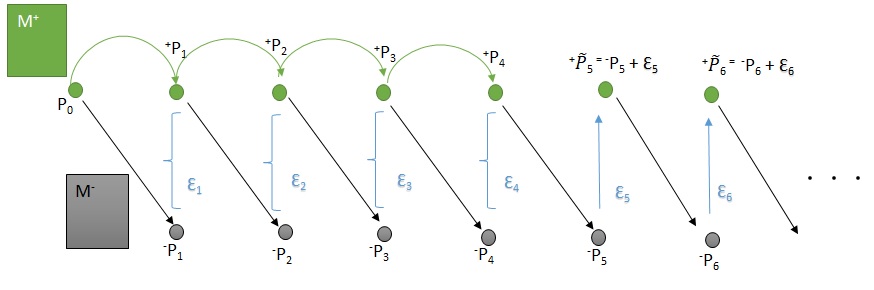}
\caption{Illustration of the stitching
 process of the {\it AI} with $r=4$ e $p=5$}
\label{fig:fig1}
\end{figure}

From the figure \ref{fig:fig1}, considering $r=4$ e $p=5$, using equations  \ref{eq42} e \ref{pesiap} one has:\\\\

$\delta^0\varepsilon_{4} = \textbf{1}\varepsilon_{4}$

$\delta^1\varepsilon_{4} = \textbf{1}\varepsilon_{4}-\textbf{1}\varepsilon_{3}$

$\delta^2\varepsilon_{4} = \textbf{1}\varepsilon_{4}-\textbf{2}\varepsilon_{3}+\textbf{1}\varepsilon_{2}$

$\delta^3\varepsilon_{4} = \textbf{1}\varepsilon_{4}-\textbf{3}\varepsilon_{3}+\textbf{3}\varepsilon_{2}-\textbf{1}\varepsilon_{1}$\\

with,\\\\
$\bar\varepsilon_{5}=\delta^0\varepsilon_{4}+\delta^1\varepsilon_{4}+\delta^2\varepsilon_{4}+\delta^3\varepsilon_{4}$.\\\\
we have:\\\\
$\bar\varepsilon_{5}=4\varepsilon_{4}-6\varepsilon_{3}+4\varepsilon_{2}-\varepsilon_{1}$ \\\\
$\bar\varepsilon_{6}=4\bar\varepsilon_{5}-6\varepsilon_{4}+4\varepsilon_{3}-\varepsilon_{2}$.\\\\

Therefore:\\\\

$P^{+}_5=P^{-}_5+4\varepsilon_{4}-6\varepsilon_{3}+4\varepsilon_{2}-\varepsilon_{1}$\\\\

$P^{+}_6=P^{-}_6+4\varepsilon_{5}-6\varepsilon_{4}+4\varepsilon_{3}-\varepsilon_{2}$.\\\\
\begin{obs}
Using the approximation (\ref{eq64}) we can  (using the mapping $M^{-}$ abd the fucntoins $\Delta^r \varepsilon_i$) calculate the point ${x_i}_{(P^{+}_{p+1})}$ with precision given by $O({\delta t}^{r})$ which is better than the precision provided by the mapping $M^{-}$.
\end{obs}

\begin{obs}
Considering the number of digits used to represent the points on the solution-curve for the system and the number  $p$ of points used to calculate the functions $\Delta^r \varepsilon_i$, we can even, em theory, reach the same level of precision (approximately) as the one obtained by using the mapping $M^{+}$.
\end{obs}

\medskip

\section{Actual implementation of the Associative Integration Algorithm - AI}
\label{actual}

In this section, we are going to present an implementation we have produced of the ideas so far introduced. So, we are going to  implement an actual version of a Associative Integration Algorithm - {\it A.I.} in a concrete setting. In the present case, we will work with the case where the mappings $M^{+}$ and $M^{-}$ are both Runge-Kutta algorithms.

We are going to do that as an implementation in C++, more specifically,  with the associated algorithms being: $RK4$ playing the role of $M^{+}$ and $RK78$ the one of $M^{-}$.

\subsection{Description of a particular implementation}

Our first objective here is to demonstrate that our idea for the {\it A.I.} works. What do we mean by ``work''? {\bf Ideally}, that means that the implementation would present the time expenditure of the mapping $M^{-}$ ($RK4$), very closely, presenting the precision of the mapping $M^{+}$ ($RK78$), again very closely. More realistically, we hope that the precision obtained will be better than the one obtainable by applying $M^{-}$ (worse than the one in the case of $M^{+}$) with the running time very close to the one of $M^{-}$. We will discuss this further along the paper.

We will call the precision of the more precise mapping $n^+$ and the one for the less precis ($M^-$), $n^-$. For the example we will display, we chose the well know Lorenz System, with a step given by $h=0.001$ chosen in order to  optimize the relation truncation/round-off errors.

To exemplify the use of the algorithm, we use a low-dimensional dynamic system: The Lorenz \cite{bib:Boyce} system, which is a chaotic dissipative system and one of the simplest that exists. The Lorenz system is given by the equations:

\begin{eqnarray}
\label{sistLorenz}
\dot{x} & = & \sigma (y-x), \nonumber \\
\dot{y} & = & - y - x z + R x,  \; \\
\dot{z} & = & x y - b z, \nonumber
\end{eqnarray}

\noindent where $ \sigma $, $ R $ and $ b $ are constant parameters of the problem (with chaotic behavior for $ R> 24.74 $). Why one of the simplest? This system has the minimum number of autonomous differential equations (time does not appear explicitly) so that there are chaos: three (in two dimensions there is no chaos, since the trajectories can not cross). Further, since chaos is a phenomenon that manifests itself only in non-linear systems, the ``least portion'' of nonlinearity we might think of adding to a linear system to make it chaotic would be a quadratic term . We can note then that the Lorenz system has only two quadratic non-linear terms.
In this implementation the values of the constants are: 
\\\\
$ b = 2.666666666666667 $ \\
$ \sigma = 10 $ \\
$ R = 28 $ \\
The initial condition used was $ (5.5,5) $.
\\

Due to the architecture of the computer we have used to run the program that implemented the {\it A.I.}, we have the following limitation for the representation of numbers: the number of digits of precision for a double-precision variable in C++ (in a 64 bits machine) is 16 decimal places at most.

This state of affairs let us to make the choices we have just mentioned: We have chosen ($RK4$) as  $M^{-}$ since its precision $n^-$ is of the order $h^4$, in our case it implies $10^{-12}$, within the precision allowed by the machine at hand. For $M^{+}$, we have chosen $RK78$ since its precision $n^+$ is of the order $h^7$, i.e., $10^{-21}$. That means that the 16 decimal places would be achieved in most cases. Since each difference $\varepsilon_i(\vec{x})$ used by {\it A.I.} provides a correction of order $h$, according to the remark \ref{ordin}, we have that for the association of the algorithms $RK4$ and $RK78$, with three levels of difference ($r=3$), would provide an improved precision of (at maximum) $h^3$, i.e., $10^{-9}$ that is within the limitation of 16 decimal places just mentioned.

In this implementation, we start off the procedure from 4 points calculated by  $M^+$ ($RK78$) and five points  calculated by  $M^-$ ($RK4$), then the  {\it A.I.} procedure calculates the fifth point (improved) related to the  $M^+$ following the stitching process that basically is made of the adding up of differences described on the remark  \ref{ordin} to correct the point given by the mapping  $M^-$.

Before embarking on showing the results that were produce by all these procedures and discuss them, we would like to present a last comment regarding all the choices mentioned above. We have chosen to work with computers with the limitation on precision mentioned above instead of going to more powerful machines and also we have chosen not to tackle the number of digits limitation by, for example, using GMP, or other forms of having arbitrary precision. The reasons for that can be divided in two aspects. We wanted to apply our ideas to a case that were more broadly applicable, to the average user of computers and numeric integration ``costumers''. Apart from that, if it works on the present scenario, it will more likely work on a more ``pure'' and powerful setting.

\begin{figure}[H]
\includegraphics[scale=0.45,bb=0 0 3 3]{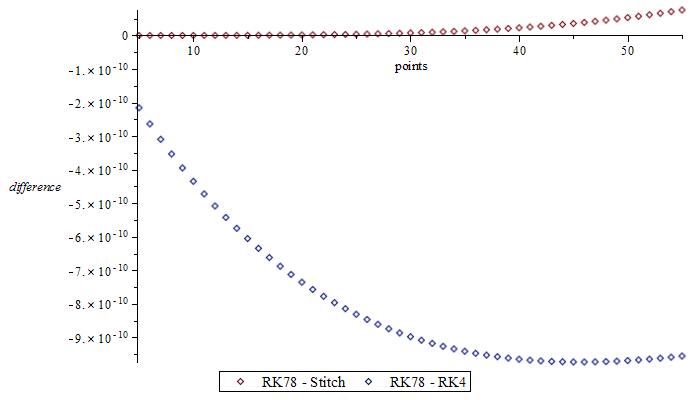}
\caption{Differences between the points of the integrators in the variable $x$}
\label{graf:grafx}
\end{figure}

\begin{figure}[H]
\includegraphics[scale=0.45,bb=0 0 3 3]{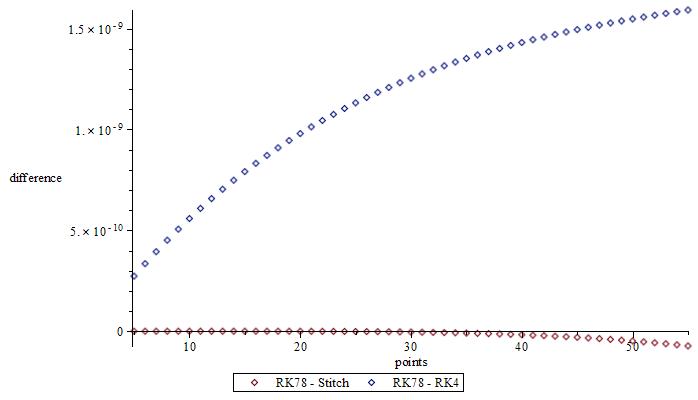}
\caption{Differences between the points of the integrators in the variable $y$}
\label{graf:grafy}
\end{figure}

\begin{figure}[H]
\includegraphics[scale=0.45,bb=0 0 3 3]{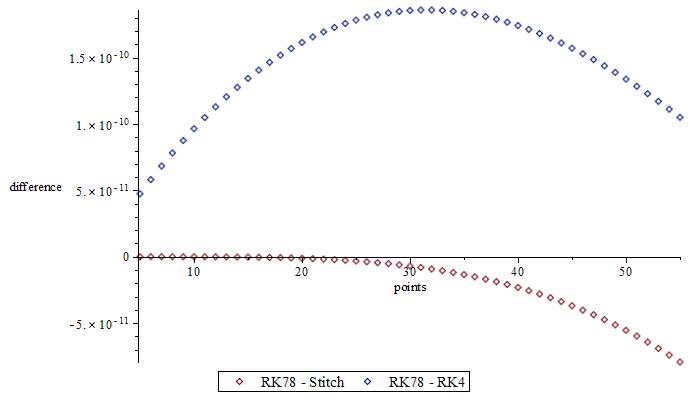}
\caption{Differences between the points of the integrators in the variable $z$}
\label{graf:grafz}
\end{figure}

From the results above, one can notice that our algorithm {\it A.I.} calculates the first point displayed with a difference, in relation to the mapping $M^+$ ($RK78$), in one decimal place only for the $x$ variable, two for the variable  $y$ and none for the variable  $Z$. On the other hand, the mapping $M^-$ ($RK4$) calculates (for the same variables) with 6 digits of difference for $x$, six for $y$ and six for  $z$. We can also observe that, for the eleventh point, our algorithm  {\it A.I.} with four decimal places of difference for the variable $x$,three for $y$ and four for the variable  $z$. Again, the mapping $M^-$ ($RK4$) has a much worse result; with six digits of difference for the variable  $x$, also six for $y$ and $z$.
The graphical display of these results is shown in figures  \ref{graf:grafx}, \ref{graf:grafy} and \ref{graf:grafz}, where the results for the differences for  $RK78$ and  $RK4$ and for $RK78$ and {\it A.I.}, again for the variables  $x$, $y$ and $z$ for the Lorenz system, are shown. As it can easily be seen by tables  (\ref{tempoRK78}), (\ref{tempoAI}) and (\ref{tempoRK4}), the running time for $RK4$ and {\it A.I.} are very close and much smaller than the one for $RK78$. This goes according to our plans so far.
On the figure \ref{graf:tdesempenho} below, we display the curves points versus running times for the three integrators: $RK4$, $RK78$ and {\it A.I.}.

\bigskip\bigskip\bigskip\bigskip\bigskip\
\\\\\\\\\\\\\\\\\


\bigskip\bigskip\bigskip\bigskip\bigskip\
\\\\\\\\\\\\\\\\\


\bigskip\bigskip\bigskip\bigskip\bigskip\
\\\\\\\\\\\\\\\\\
\

\begin{table}[b]
\caption{Performance of RK78}
\label{tempoRK78}
\begin{tabular}{c|c}
\hline
Points & Time (s)\\
\hline
\hspace{2cm}5\hspace{2cm}  & \hspace{2cm}0,000034\hspace{2cm} \\
10 & 0,000055 \\
15 & 0,000076 \\
20 & 0,000097 \\
25 & 0,000118 \\
30 & 0,000138 \\
35 & 0,000160 \\
40 & 0,000181 \\
45 & 0,000201 \\
50 & 0,000223 \\
55 & 0,000242 \\
\hline
\end{tabular}
\end{table}
~\

\begin{table}[t]
 \caption{The Stitching process performance }
\label{tempoAI}
\begin{tabular}{c c}
\hline
Pontos & Tempo (s)\\
\hline
\hspace{2cm}5\hspace{2cm}  &\hspace{2cm} 0,000017\hspace{2cm} \\
10 & 0,000022 \\
15 & 0,000026 \\
20 & 0,000032 \\
25 & 0,000036 \\
30 & 0,000044 \\
35 & 0,000046 \\
40 & 0,000051 \\
45 & 0,000056 \\
50 & 0,000061 \\
55 & 0,000066 \\
\hline
\end{tabular}
\end{table}

\begin{table}[t]
\caption{Performance of RK4}
\label{tempoRK4}
\begin{tabular}{c c}
\hline
Points & Time (s)\\
\hline
\hspace{2cm}5\hspace{2cm}  &\hspace{2cm} 0,000020\hspace{2cm} \\
10 & 0,000024 \\
15 & 0,000028 \\
20 & 0,000032 \\
25 & 0,000036  \\
30 & 0,000040 \\
35 & 0,000045 \\
40 & 0,000048 \\
45 & 0,000052 \\
50 & 0,000057 \\
55 & 0,000060 \\
\hline
\end{tabular}
\end{table}


Through the graph we observe the linear behavior of the variation of the number of points calculated as a function of time, which was expected because the times are related to the number of operations performed by each integrator in the numerical integration process. The angular coefficients of the lines of each integrator represent the rate of change of the number of points as a function of time, that is, of the integrator's performance over time. Therefore the higher the slope of the line the better the performance of the integrator.

As seen in the previous tables the RK4 times were very close to the times of {\it AI}, a fact that is very clear when we compare the angular coefficients of the RK4 and the {\it AI} lines. Already the angular coefficient of the line of RK78 is smaller than the angular coefficient of the line of {\it AI}.
The ratio between the angular coefficients $\alpha_{ {\it AI}}$ and the  $\alpha_{Rk78}$ provides the time constant relative to the improvement in the efficiency of the process.
Being $\alpha_{{\it AI}} = 1.10^6$ and $\alpha_{Rk78} = 239640$ we have:
\begin{equation}
\frac{\alpha_{ {\it AI}}}{\alpha_{Rk78}} \simeq 4.
\end {equation}
This means that the stitching process of {\it AI} is approximately 4 times faster than RK78.
However, at this stage, the accuracy of {\it AI} decreases after the calculation of some points as seen in the figures \ref{graf:grafx}, \ref{graf:grafy} and \ref{graf:grafz}. Therefore, to obtain an associative integrator with the ability to calculate a large number of points, it is necessary to cease the stitching process (after checking the number of points with the desired precision), recalculate the amount of $ p-1 $ points with the mapping $ M ^ + $, recalculate the amount of $ p $ points with the mapping $ M ^ - $, having as initial condition the mapping points $ M ^ + $ to restart the stitching process, repeating these steps until you reach the total number of points you want.

\section{Performance of {\it AI}}
\label{performance AI}
As we saw at the end of the previous section, although at the beginning of the process the calculation of points using {\it AI} is exciting (with an accuracy that equals the accuracy of the mapping $ M ^ + $ (RK78)), accuracy drops after some points. As the interest is to build a stable integrator, we implement the following process (taking the Lorenz system as a model):
\begin {enumerate}
\item Choose $ p = 4 $ and $ r = 3 $.
\item Make an integration of 10 points:
\begin {itemize}
\item 1 point - initial condition;
\item 3 points integrated with RK78;
\item 4 points integrated (1 from the initial condition and 3 from the points calculated with the RK78) with the RK4 to do the ``stitching'';
\item 7 points with the ``stitching'' of {\it AI}, thus obtaining point 11.
\end {itemize}
\item From the obtained point 11, repeat the process described in item 2, that is, calculate another 3 points with the $ M ^ + $ (RK78) (points 12, 13 and 14), 4 points with the $ M ^ - $ (RK4) for ``sewing'' again and 7 more stitches, with the ``stitching'' of {\it AI}, (from 15 to 21).
\item Repeat the procedure described in item 2 until you reach the desired number of points, that is, make 3 more points with the $ M ^ + $ (RK78), plus 4 points with $ M ^ - $ (RK4) to make the `stitching ', 7 stitches with the ``stitching'' of {\it AI} and so on.
\end {enumerate}

Since $T_{RK78} \simeq 4T_{{\it AI_{stitch}}}$ and $T_{RK4} \simeq T_{{\it AI_{stitch}}}$ is the result of the calculation of 10 points with the $ AI $, we have the following performance for the calculation of 10 points:

\begin {itemize}

\item Time for the RK78 to calculate 10 points in relation to the time of {\it AI}, \\

$ T_RK78 = 10.4 = 40 $ (arbitrary unit) \\\\

\item Total time of {\it AI} taking into account the 3 steps of RK78 and 4 steps of RK4. \\

$ T_{it AI} = 3.4 + 4.1 + 7.1 = 23 $ (arbitrary unit) \\
\end {itemize}

\bigskip

\begin{figure}[b]

\includegraphics[scale=0.45,bb=0 0 3 3]{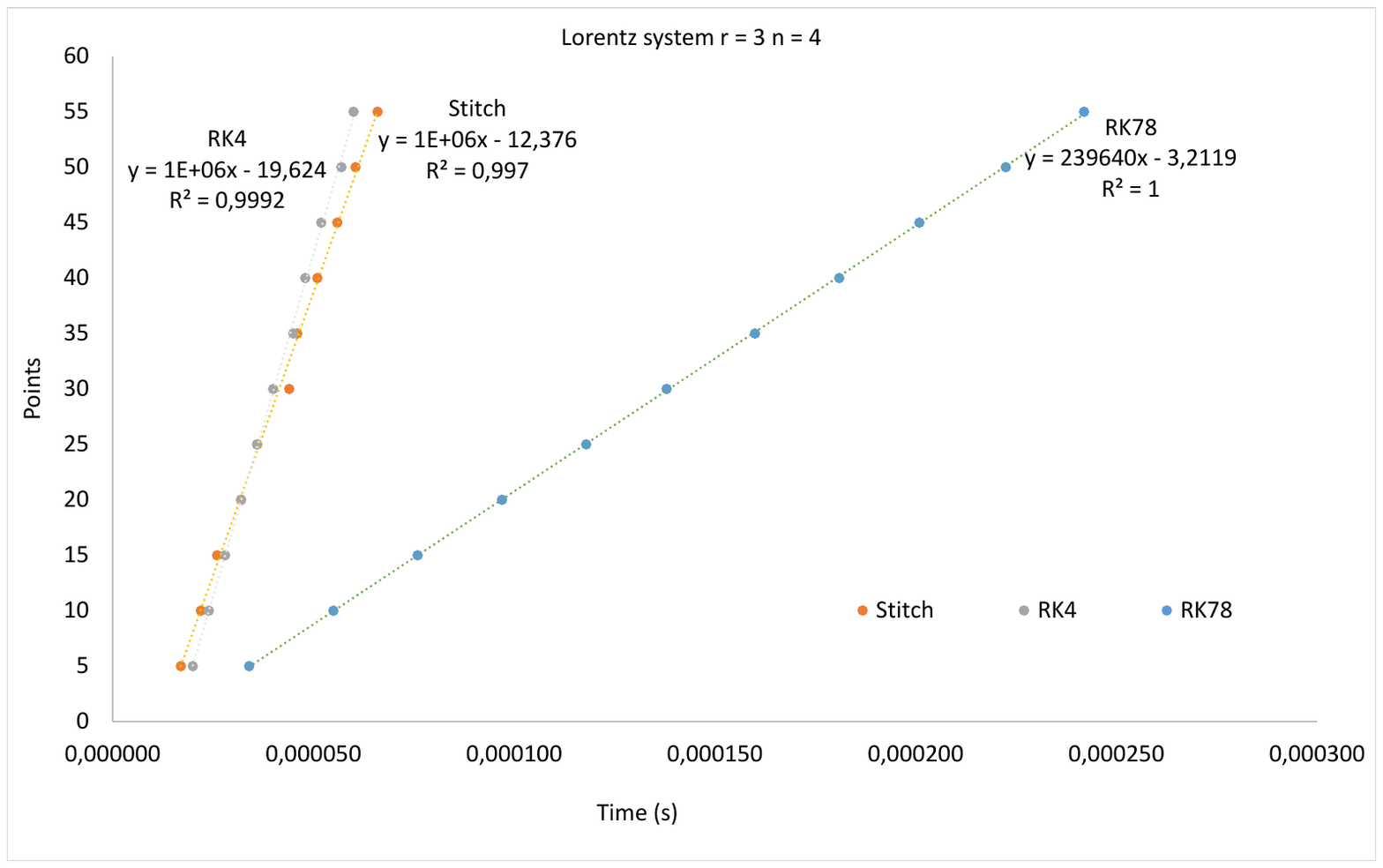}
\caption{running time performance for the  integrators}
\label{graf:tdesempenho}
\end{figure}
\FloatBarrier

\begin{figure}[h]
\includegraphics[scale=0.45,bb=0 0 3 3]{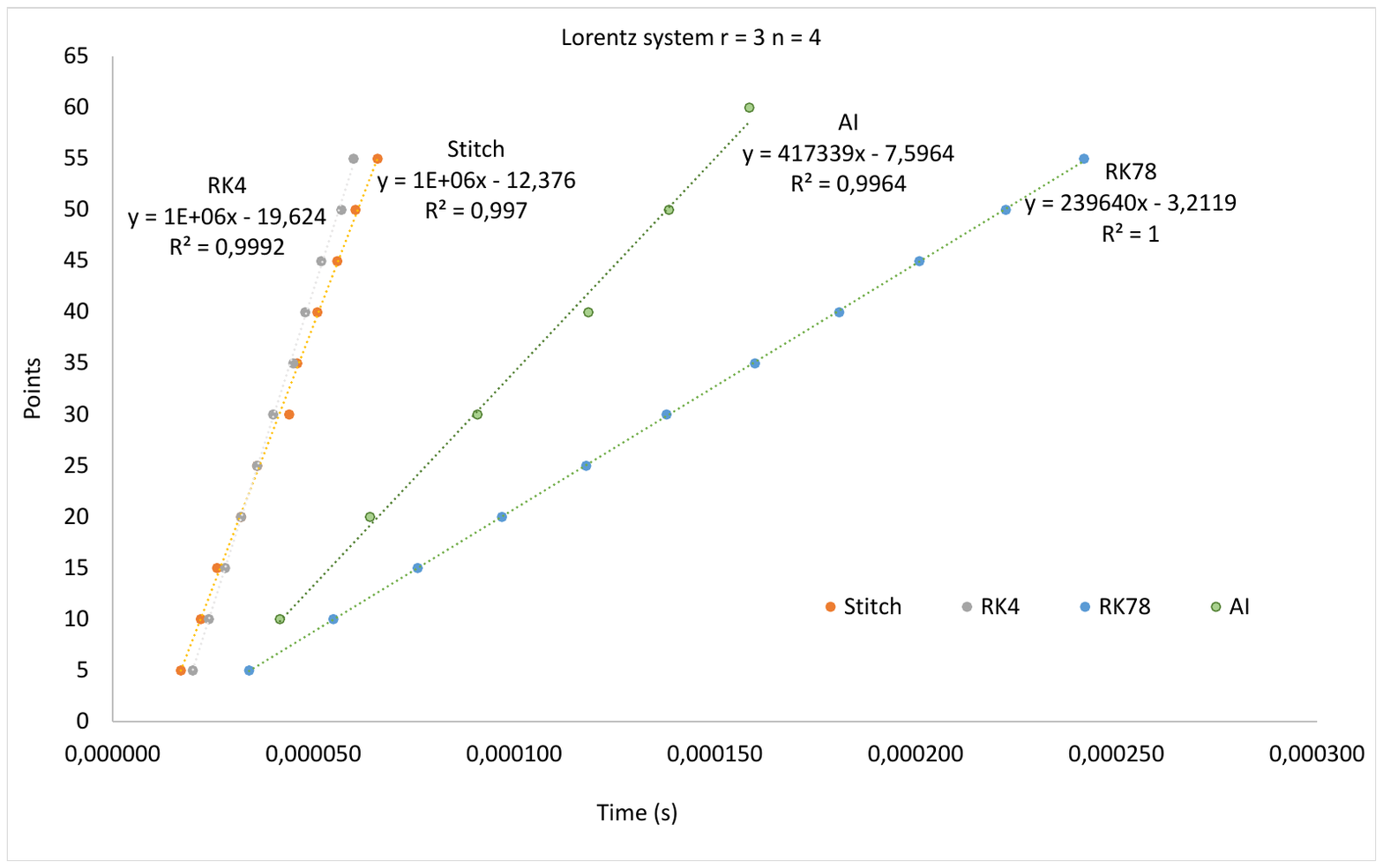}
\caption{running time performance for the integrators}
\label{graf:tdesempenho2}
\end{figure}

\begin{figure}[h]
\includegraphics[scale=0.53,bb=0 0 3 3]{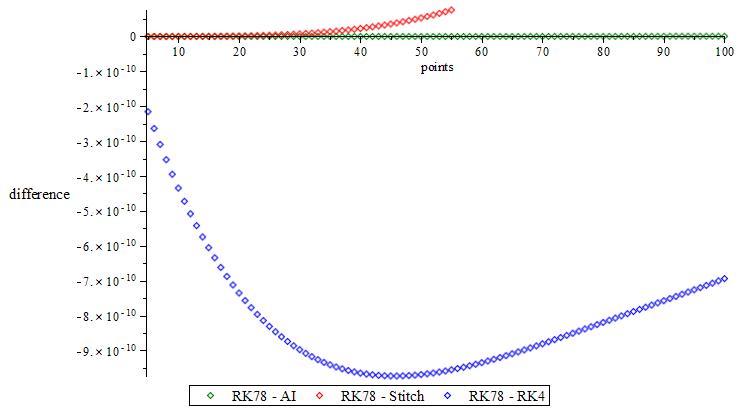}
\caption{Difference between
 RK78 and RK4, RK78 and AI, and RK78 and stitching
, coordinate $x$.}
\label{graf:grafdifx}
\end{figure}

\begin{figure}[h]
\includegraphics[scale=0.50,bb=0 0 3 3]{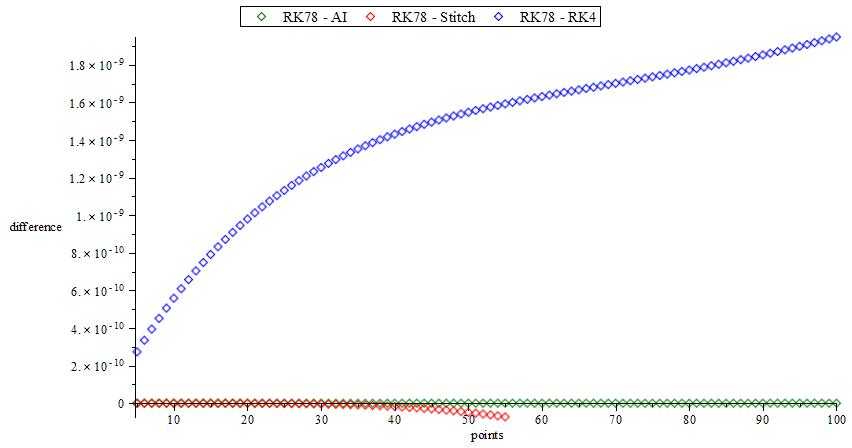}
\caption{Difference between
 RK78 and RK4, RK78 and AI, and RK78 and stitch
, coordinate $y$.}
\label{graf:grafdify}
\end{figure}

\begin{figure}[h]
\includegraphics[scale=0.50,bb=0 0 3 3]{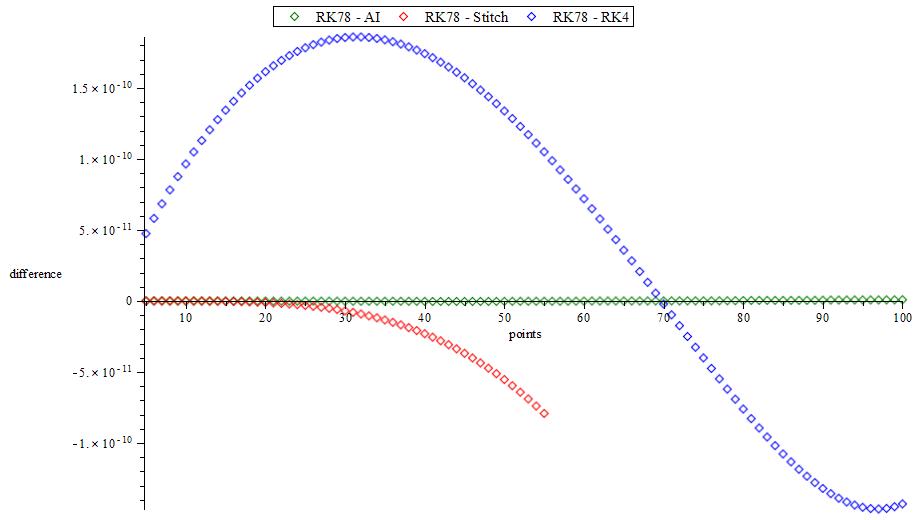}
\caption{Difference between
 RK78 and RK4, RK78 and AI, and RK78 and stitching, coordinate $z$.}
\label{graf:grafdifz}
\end{figure}

That is, under these conditions, we obtain a resultant associative integrator more accurate than RK4, with a precision close to RK78 and with a time performance approximately half the time of RK78, confirming the viability of the associative integration process.
The graph \ref{graf:tdesempenho2} shows the time performance.
The \ref {graf:grafdifx}, \ref {graf:grafdify} and \ref {graf:grafdifz} charts show the precision of this resulting associative integrator for the Lorenz system.

\bigskip\
\\\\\\\\\\\\\\\\\\\\\\\\\\\\\\

\bigskip\
\\\\\\\\\\\\\\\\\\\\\\\\\\\\\\\\\\\\\\\\

\FloatBarrier

\bigskip\
\\\\\\\\\\\\\\\\\\\\\\\\\\\\\\\\\

\bigskip\
\\\\\\\\\\\\\\\\\\\\\\\\\\\\\\\

\section {Other examples of systems calculating with {\it AI}}
\begin{itemize}
\item Rössler attractor

Another example of the use of the algorithm will be the Rössler attractor. That is also a chaotic system of low dimensionality \cite{letellier2010influences}.
The Rössler system is given by the equations:
.
\begin{eqnarray}
\label{rosslersistem}
\dot{x} & = & -y -x, \nonumber \\
\dot{y} & = &  x +  ay,  \; \\
\dot{z} & = & b + z(x - c), \nonumber
\end{eqnarray}

\noindent where $ \ a $, $ b $ and $ c $ are constant parameters of the problem ($a=0.2$, $b=0.2$ and $c=5.7$).
The initial condition used was $ (0.1,0.1,0.1)$.
Parameters used in {\it AI}: $h=0.008$, $p=4$ and $r=3$.
The running time performance and the accuracy of the AI are shown in the graph \ref{graf:rosslerX}, \ref{graf:rosslerY}, \ref{graf:rosslerZ} and table \ref{temporossler}.

~\\\\\\\\\\\\\\\\\\\\\\\\\\\\

  \begin{figure}[h]
\includegraphics[scale=0.50,bb=0 0 3 3]{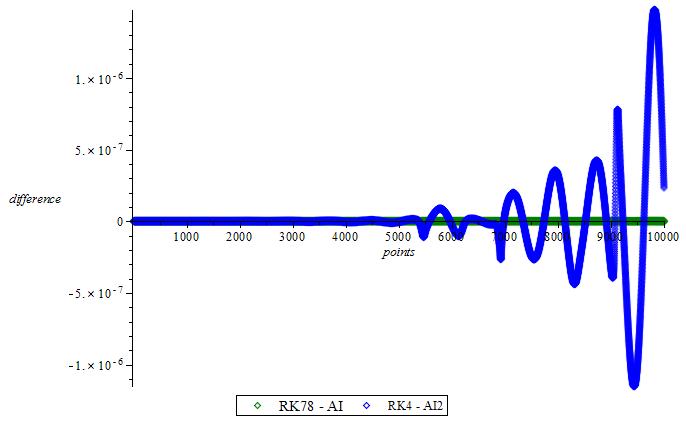}
\caption{Difference between
 RK78 and RK4, RK78 and AI, coordinate $x$ for Rössler system.}
\label{graf:rosslerX}
\end{figure}

Since the scale of the graph does not allow us to observe the difference in the whole range of 10,000 points we have the graphs \ref{graf:rosslerx1000} and \ref{graf:rosslerx5000} to show this difference in the appropriate scale for the 
coordinate y.
  ~\\\\\\\\\\\\\\\\\\\\\\\\\\\\\
  
\begin{figure}[H]
\includegraphics[scale=0.50,bb=0 0 3 3]{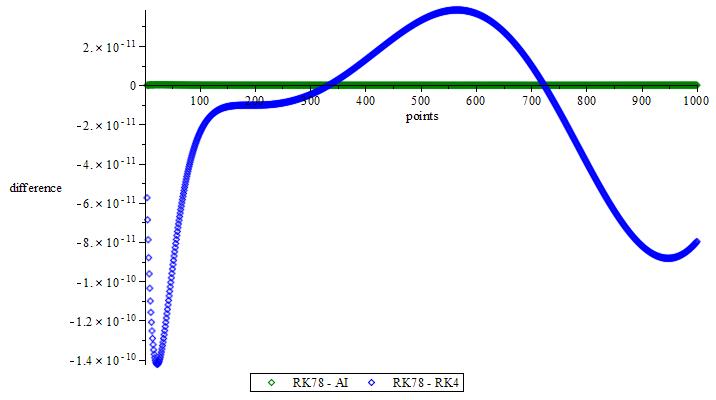}
\caption{Difference between
 RK78 and RK4, RK78 and AI, coordinate $x$ from 0 to 1000 points for Rössler system.}
\label{graf:rosslerx1000}
\end{figure}
\FloatBarrier
 ~\\\\\\\\\\\\\\\\\\\\\\\\\\\\\

 \begin{figure}[h]
\includegraphics[scale=0.50,bb=0 0 3 3]{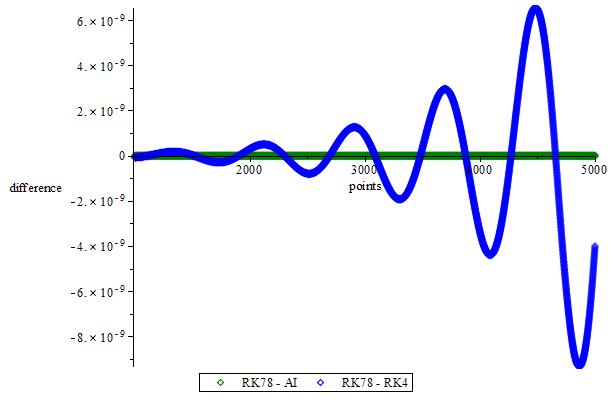}
\caption{Difference between
 RK78 and RK4, RK78 and AI, coordinate $x$ from 1000 to 5000 points for Rössler system.}
\label{graf:rosslerx5000}
\end{figure}

 ~\\\\\\\\\\\\\\\\\\\\\\\\\\
 \begin{figure}[H]
\includegraphics[scale=0.40,bb=0 0 3 3]{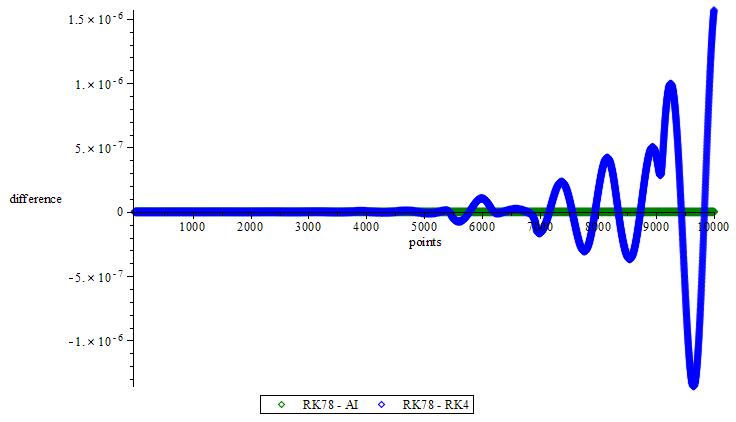}
\caption{Difference between
 RK78 and RK4, RK78 and AI, coordinate $y$ for Rössler system.}
\label{graf:rosslerY}
\end{figure}
 \begin{figure}[b]
\includegraphics[scale=0.50,bb=0 0 3 3]{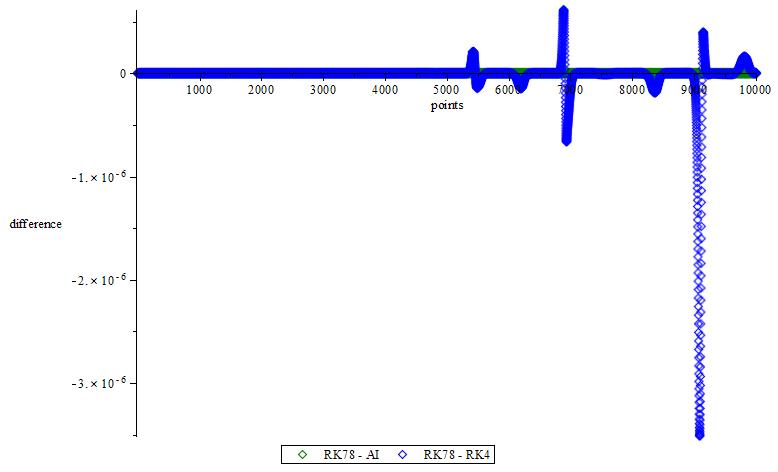}
\caption{Difference between
 RK78 and RK4, RK78 and AI, coordinate $z$ for Rössler system.}
\label{graf:rosslerZ}
\end{figure}
~\\\\\\\\\\\\\
\begin{table}[H]
\caption{Integrators time to 10000 points for Rössler system }
\label{temporossler}
\begin{tabular}{c c}
\hline
integrator & Time (s)\\
\hline
\hspace{2cm}\hspace{2cm}  &\hspace{2cm} \hspace{2cm}\hspace{2cm}\hspace{2cm}\\
RK4 & 0,008128 \\
RK78 & 0,036718 \\
AI & 0,019093 \\

\hline
\end{tabular}
\end{table}

~\\\\\\\\\\\\\\\\\\\\\\\\\\\\\\\\\\\\
\item T System 

The T system is a nonlinear 3D chaotic system derived from a variant of the Lorentz system proposed to study causal conditions that allows a greater possibility in the choice of system parameters and therefore exhibits a more complex dynamics. \cite{tigan2008analysis}.
The T system is given by the equations:

\begin{eqnarray}
\label{sisT}
\dot{x} & = & a(y -x), \nonumber \\
\dot{y} & = &  (c -a)x - axy,  \; \\
\dot{z} & = & bz + xy, \nonumber
\end{eqnarray}

\noindent where $ \ a $, $ b $ and $ c $ are constant parameters of the problem ($a=2.1$, $b=0.6$ and $c=30$).
The initial condition used was $ (0.1,-0.3,0.2)$.
Parameters used in {\it AI}: $h=0.001$, $p=4$ and $r=3$.
The running time performance and the accuracy of the AI are shown in the graph \ref{graf:IX}, \ref{graf:IY}, \ref{graf:IZ} and table \ref{tempoT}.

~\\\\\\\\\\\\\\\\\\\\\\\\\\\\\\\\\\\\\\\\\\\\\\\\\\\\\

  \begin{figure}[h]
\includegraphics[scale=0.50,bb=0 0 3 3]{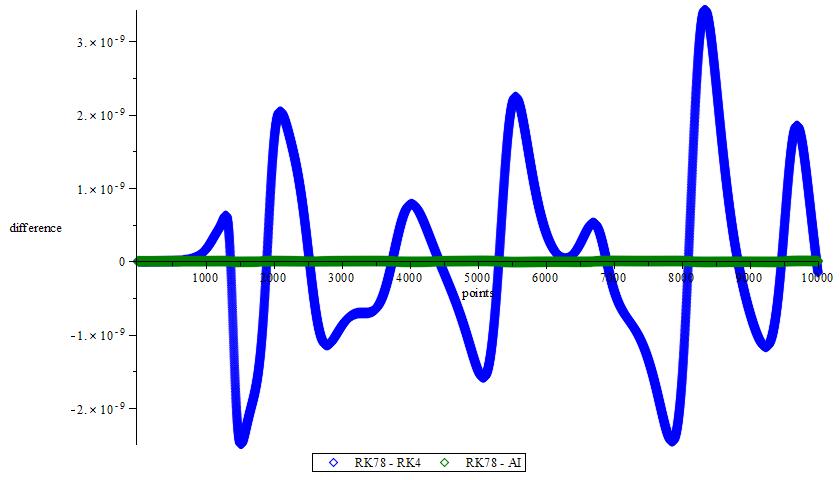}
\caption{Difference between
 RK78 and RK4, RK78 and AI, coordinate  $x$ for T system.}
\label{graf:IX}
\end{figure}

  ~\\\\\\\\\\\\\\\\\\\\\\\\\\\\\
  
  \begin{figure}[H]
\includegraphics[scale=0.50,bb=0 0 3 3]{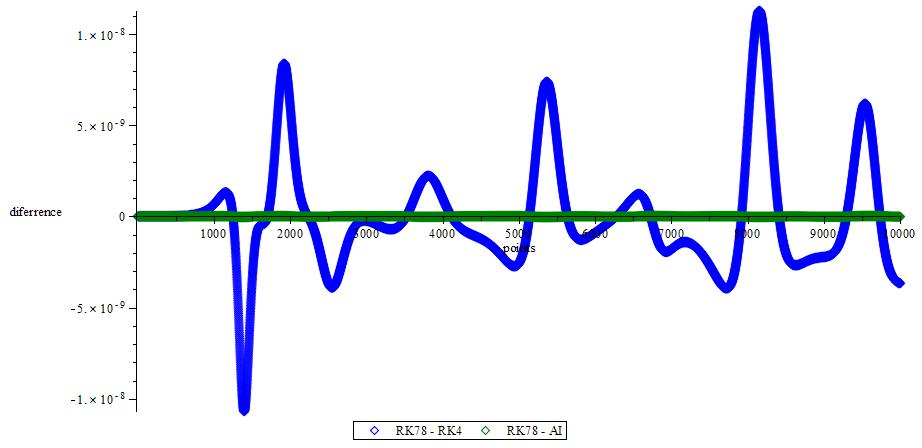}
\caption{Difference between
 RK78 and RK4, RK78 and AI, coordinate $y$ for T system.}
\label{graf:IY}
\end{figure}

~\\\\\\\\\\\\\\\\\\\\\\\\\\\\\\\\\\\\\\\\\

\begin{figure}[h]
\includegraphics[scale=0.50,bb=0 0 3 3]{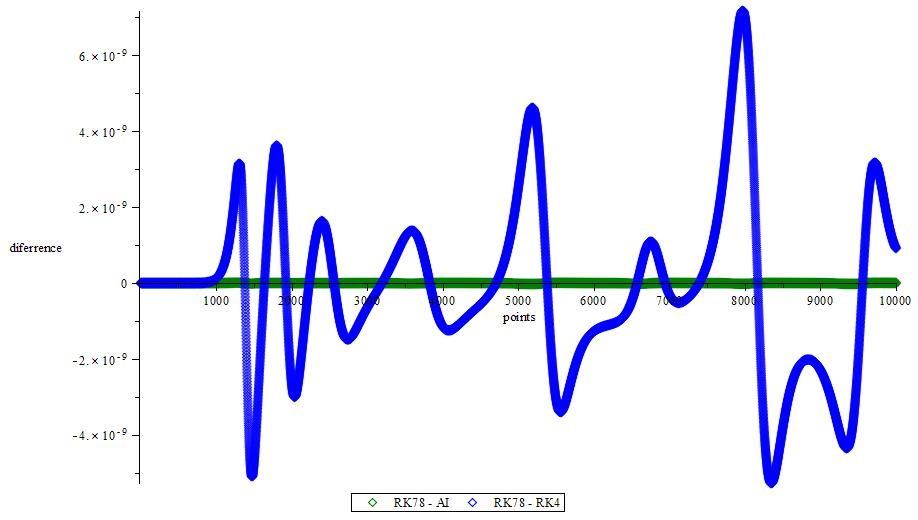}
\caption{Difference between
 RK78 and RK4, RK78 and AI, coordinate $Z$ for T system.}
\label{graf:IZ}
\end{figure}

\begin{table}[H]
\caption{Integrators time to 10000 points for T system}
\label{tempoT}
\begin{tabular}{c c}
\hline
integrator & Time (s)\\
\hline
\hspace{2cm}\hspace{2cm}  &\hspace{2cm} \hspace{2cm}\hspace{2cm}\hspace{2cm}\\
RK4 & 0,008353 \\
RK78 & 0,037705 \\
AI & 0,018950 \\
\hline
\end{tabular}
\end{table}
~\\\\\\\\\\\\\\\\\\\

\item The Hénon-Heiles systems

The Hénon-Heiles model was initially created to describe the stellar movement \cite{henon1964applicability}. It also describes the motion of molecules coupled in a non-linear fashion \cite{barrow2003test}. Currently, this conservative system is the object of much study in the area of analysis of nonlinear  systems \cite{letellier2005relation}, \cite{aguirre2008observability}. The Hénon-Heiles system can be described by the following set of four ordinary differential equations:

\begin{eqnarray}
\label{sisTww}
\dot{x} & = & u, \nonumber \\
\dot{y} & = &  v,  \; \\
\dot{u} & = &  -x - 2xy,  \nonumber \\
\dot{v} & = & -x^2 + y^2 - y, \nonumber
\end{eqnarray}

The initial condition used was $ (0.000, 0.670, 0.093,0.000 ) $.
Parameters used in {\it AI}: $h=0.007$, $p=4$ and $r=3$.
The running time performance and the accuracy of the AI are shown in the graph 
\ref{graf:henonx}, \ref{graf:henony}, \ref{graf:henonu} and table \ref{graf:henonv}.
\\\\\\\\\\\\\\\\\\\\\\\\\\\\\\\\\\\\\\\\\\\\\\\\\\\\\\\\
\begin{figure}[h]
\includegraphics[scale=0.50,bb=0 0 3 3]{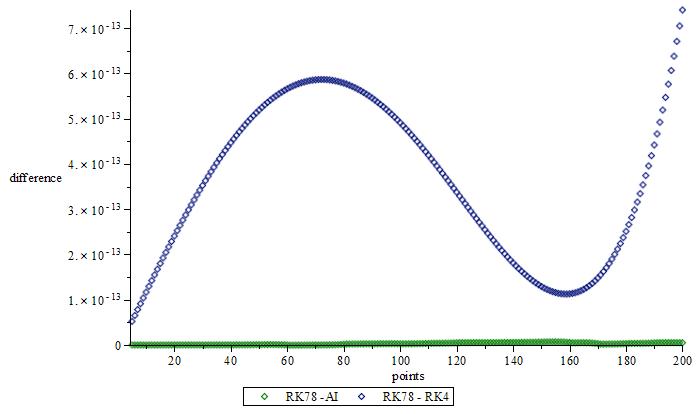}
\caption{Difference between
 RK78 and RK4, RK78 and AI, coordinate  $x$ for Hénon-Heiles system.}
\label{graf:henonx}
\end{figure}
\\\\\\\\\\\\\\\\\\\\\\\\\\\\\\\\\
\begin{figure}[h]
\includegraphics[scale=0.50,bb=0 0 3 3]{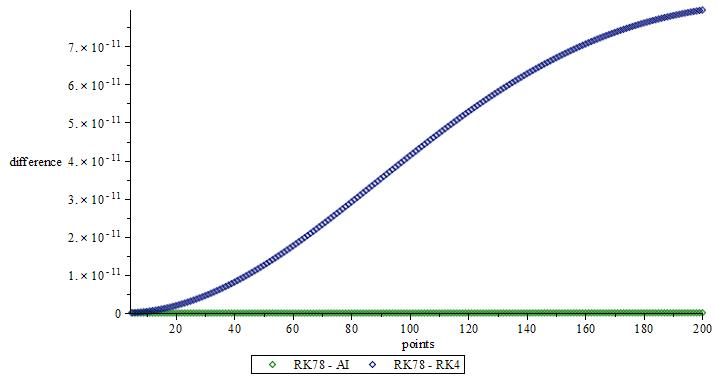}
\caption{Difference between RK78 and RK4, RK78 and AI, coordinate  $y$ for Hénon-Heiles system.}
\label{graf:henony}
\end{figure}
\\\\\\\\\\\\\\\\\\\\\\\\\\\\\\\\\\\
\begin{figure}[h]
\includegraphics[scale=0.50,bb=0 0 3 3]{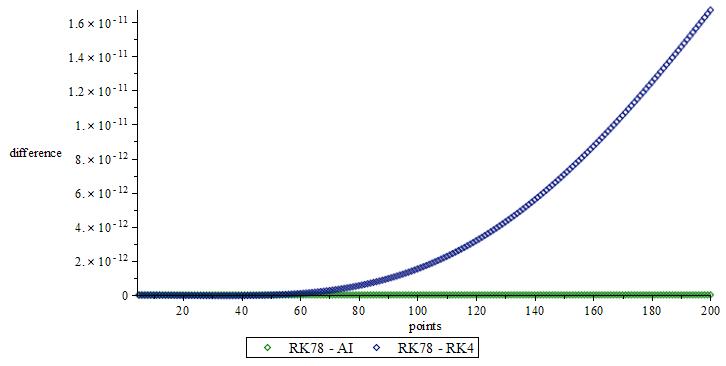}
\caption{Difference between
 RK78 and RK4, RK78 and AI, coordinate  $u$ for Hénon-Heiles system.}
\label{graf:henonu}
\end{figure}
\\\\\\\\\
\begin{table}[b]
\caption{Integrators time to 1000 points for Hénon-Heiles system}
\label{tempohenon}
\begin{tabular}{c c}
\hline
integrator & Time (s)\\
\hline
\hspace{2cm}\hspace{2cm}  &\hspace{2cm} \hspace{2cm}\hspace{2cm}\hspace{2cm}\\
RK4 & 0,001314 \\
RK78 & 0,005620 \\
AI & 0,003093 \\
\hline
\end{tabular}
\end{table}
~\\\\\\\\\\\\\\\\\\\\\\\\\\\\\\\\
\begin{figure}[h]
\includegraphics[scale=0.50,bb=0 0 3 3]{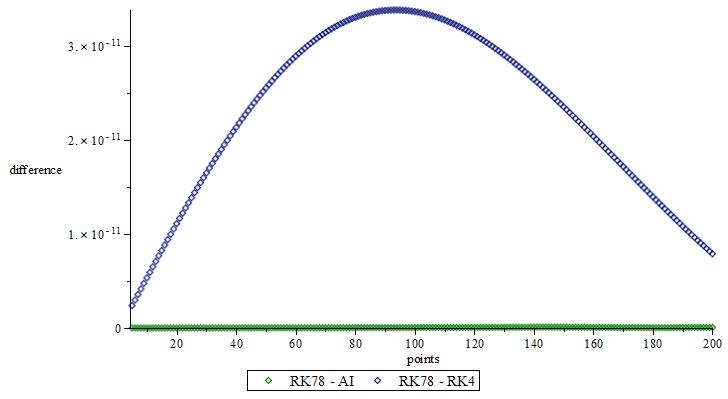}
\caption{Difference between
 RK78 and RK4, RK78 and AI, coordinate  $v$ for Hénon-Heiles system.}
\label{graf:henonv}
\end{figure}

\end{itemize}
 {\bf \large{The steps of the Algorithm (Pseudo code):}}

\begin{enumerate}

\item Choose two mappings, $M^{+}$ e $M^{-}$, with different levels of precision.

\item Chose an initial point $P_0$, a number of points N,  and the total number of points  $Npoints$ to be calculated.

\item Choose a positive integer  $p$ and, using the mapping with higher precision, $M^{+}$, calculate  $p$ points after the initial condition $P_0$ (the points $P^{+}$).

\item From those  $p+1$ points, calculate, using the mapping with the lesser precision, $M^{-}$, $\,p+1$ points from the  $p+1$ points $\{P_0,\,P^{+}_1,\ldots,\,P^{+}_p\}$ (the points $P^{-}$).

\item Using the set of points $\{P_0\} \cup \{P^{+}\} \cup \{P^{-}\}$ calculate the functions $\Delta^r \varepsilon_i$, $(r=0,\ldots,p)$.

\item Calculate the corrected point ${x_i}_{(P^{+}_{p+1})}$ using the relation ${x_i}_{(P^{+}_{p+1})} \approx {x_i}_{(P^{-}_{p+1})} +\sum_{k=0}^{r}\Delta^k\varepsilon_i(\vec{x})_{(P^+_{p+1})}$.

\item Calculate the point ${x_i}_{(P^{-}_{p+2})}$ using the mapping $M^{-}$ on the corrected point ${x_i}_{(P^{+}_{p+1})}$.

\item Calculate the functions $\Delta^r \varepsilon_i$, $r=(0,\ldots,p)$ for the point ${x_i}_{(P^{+}_{p+1})}$.

\item Calculate the corrected value for the point ${x_i}_{(P^{+}_{p+2})}$ using the relation ${x_i}_{(P^{+}_{p+2})} \approx {x_i}_{(P^{-}_{p+2})} +\sum_{k=0}^{r}\Delta^k\varepsilon_i(\vec{x})_{(P^+_{p+2})} $, up to points N.

\item Repeat steps 4, 5, 6, 7, 8, and 9 for N at N points, using as the initial condition the last point of the previous step, up to the total number of points ${Npoints}$.

\end{enumerate}
\newpage
\section*{Conclusion}
Numerical integrators have a great relevance in the resolution of
dynamic systems. This is because most of the problems that are modeled
do not have an analytical solution, forcing researchers (in physics and 
other diverse areas) to use this technique. Making these numerical 
solutions more precise (and with lower running time) is one of the  most important challenges one has to face in the area of numerical integration. The relationship 
between precision and execution time is directly related to the degree
of complexity of the adopted integrator. The greater the order of the
integrator, the greater the number of operations that it must perform  (as seen in section (\ref{jest})), correspondingly reducing its performance. On the other hand, the simpler (lower order), the fewer the operations to be performed, which makes the integrator faster but less accurate. In order to optimize the relation {\bf processing time x precision} we have developed  an algorithm that produces a mixed integrator that, through the association of two integrators with distinct precision, generates an
integrator that approximates the precision of the most precise 
integrator in a execution time close to that of the integrator less
precise. To demonstrate that the idea of the associative integration
algorithm $(AI)$ actually works in real dynamic systems, we have produced a computational implementation and applied it to many well known chaotic systems of low 
dimensionality. In this implementation we used as integrators two 
integrators of the Runge-Kutta family, the $RK4$ and the $RK78$,
generating the $AI_{RK (4-78)}$. With this implementation of the 
integrator (in the case for the Lorentz system, with $p = 4$ and $r = 3$) it is possible to observe,through figures \ref{graf:grafx}, \ref{graf:grafy} and \ref{graf:grafz}, that for
a few points the precision of the $ AI $ stitching process approaches the
precision of the $ M ^ + $ mapping. However, from the same figures just mentioned, one can immediately notice that, after 30 pints or so, the difference of the points calculated using AI and RK78 starts to grow.Something had to be done if one wants to be able to integrated much more points ahead.

What we have done was to realize that, to
calculate a larger number of points using $AI$, it was necessary to stop the stitching process, after some points, where it still has the desirable precision, and start over the process. So, in a way, we had to make cycles of "stitching" to keep our goal of achieving the precision of the $M^+$ integrator, please see section \ref{performance AI}. 

As can be seen from figures \ref{graf:grafdifx}, \ref{graf:grafdify} and \ref{graf:grafdifz}, again for the Lorentz system, what we called in the figures ``stitching'' (the old procedure without the cycles) departures from the results obtained by the RK78 integrator around 30 points. But, for what we have called  ``AI'' on the same pictures, the results obtained via ``AI'' follow those of RK78 much longer, as desired.

We have also been able to verify, through figure \ref{graf:tdesempenho2}, that the running time relation of the ``simple `stitching'' process  to the $M^+$ mapping (RK78) is approximately 4 times faster and that it has a running time performance very close to the mapping $M^-$ (RK4). But, of course, this running time performance suffered in the case of ``AI'' (the stitching with the cycles mentioned above). Again, as can be seen in figure \ref{graf:tdesempenho2}, the resulting Associative Algorithm time will be approximately half the mapping time $M^+$, for this particular implementation, which makes it clear that the idea of the associative algorithm still works.

According to the results obtained from the comparisons of the solutions of the Rössler, T and Hénon-Heiles systems given by the integrators, we see that the precision of the  {\it {AI}} (half the time of the RK78) follows the accuracy of the RK78 with was expected. The running time performance also remained within the expected due to the fact that it is linked to the number of operations performed by each integrator, as we have theoretically expected and it was confirmed by the results in tables \ref{temporossler}, \ref{tempoT} and \ref{tempohenon}.

In addition, we saw a potential in the
algorithm that consists of using it as a precision modulator, using  different settings for number of differences ``{\bf $r$}'' and the number of initial points ``{\bf $p$}'', one can vary the precision between the precision values $n^-$ from the mapping $M^-$ and $n^+$ from the mapping $M^+$ (provided there is a freedom to 
represent the number of decimal places in the working machine). Therefore, we have a new approach to the numerical resolution of dynamic
systems capable of improving the precision of a simpler integrator in a
execution time smaller than that of a more robust integrator, generating
an optimization that can be very interesting for the user.

~\\\\\\\\\\\\\\

\section*{References}
\bibliographystyle{unsrt}
\bibliography{bibliografia}

\begin{thebibliography}{10}

\bibitem{cellier}
E.~Cellier, F.E.;~Kofman.
\newblock {\em Continuous system simulation}.
\newblock Springer Science \& Business Media, 2006.

\bibitem{drazin1992nonlinear}
P.~G Drazin.
\newblock {\em Nonlinear systems}, volume~10.
\newblock Cambridge University Press, 1992.

\bibitem{JCAM}
J.~Avellar, L.G.S. Duarte, S.E.S. Duarte, and L.A.C.P. da~Mota.
\newblock Integrating first-order differential equations with liouvillian
  solutions via quadratures: a semi-algorithmic method.
\newblock {\em Journal of Computational and Applied Mathematics}, 182:327,
  2005.

\bibitem{AMC}
J.~Avellar, L.G.S. Duarte, S.E.S. Duarte, and L.A.C.P. da~Mota.
\newblock A semi-algorithm to find elementary first order invariants of
  rational second order ordinary differential equations.
\newblock {\em Appl. Math. Comp.}, 184:2, 2007.

\bibitem{JMP}
L.G.S. Duarte and L.A.C.P .da Mota.
\newblock Integrals for rational second order ordinary differential equations.
\newblock {\em J. Math. Phys.}, 50:013514, 2009.

\bibitem{elementary3D}
L.G.S. Duarte and L.A.C.P .da Mota.
\newblock 3d polynomial dynamical systems with elementary first integrals.
\newblock {\em Journal of Physics A: Mathematical and Theoretical}, 43(6),
  2010.

\bibitem{secondTHEOps1}
L.G.S. Duarte, S.E.S. Duarte, and L.A.C.P. da~Mota.
\newblock Analyzing the structure of the integrating factors for first order
  ordinary differential equations with liouvillian functions in the solution.
\newblock {\em J. Phys. A: Math. Gen.}, 35:1001, 2002.

\bibitem{firsTHEOps1}
L.G.S. Duarte, S.E.S. Duarte, and L.A.C.P. da~Mota.
\newblock A method to tackle first order ordinary differential equations with
  liouvillian functions in the solution.
\newblock {\em . Phys. A: Math. Gen.}, 35:3899, 2002.

\bibitem{PS2}
L.G.S Duarte, S.E.S. Duarte, L.A.C.P. da~Mota, and J.F.E. Skea.
\newblock Solving second order ordinary differential equations by extending the
  prelle-singer method.

\bibitem{nossoPS1CPC}
L.G.S. Duarte, S.E.S. Duarte, L.A.C.P. da~Mota, and J.F.E. Skea.
\newblock Extension of the prelle-singer method and a maple implementation.
\newblock {\em Computer Physics Communications}, 144(1):46, 2002.

\bibitem{ndyn}
L.G.S. Duarte, H.P. Oliveira, R.O. Ramos, L.~A. C.~P. da~Mota, and J~E~F Skea.
\newblock Numerical analysis of dynamical systems and the fractal dimension of
  boundaries.
\newblock {\em Computer Physics Communications}, 119(2-3):256, 1999.

\bibitem{imp}
P.R.L. Alves, L.G.S. Duarte, and L.A.C.P. da~Mota.
\newblock Improvement in global forecast for chaotic time series.
\newblock {\em Computer Physics Communications}, 207(Supplement C):325 -- 340,
  2016.

\bibitem{new}
P.R.L. Alves, L.G.S. Duarte, and L.A.C.P. da~Mota.
\newblock A new method for improved global mapping forecast.
\newblock {\em Computer Physics Communications}, 207(Supplement C):539 -- 541,
  2016.

\bibitem{alt}
P.R.L. Alves, L.G.S. Duarte, and L.A.C.P. da~Mota.
\newblock Alternative predictors in chaotic time series.
\newblock {\em Computer Physics Communications}, 215(Supplement C):265 -- 268,
  2017.

\bibitem{nc}
P.R.L. Alves, L.G.S. Duarte, and L.A.C.P. da~Mota.
\newblock A new characterization of chaos from a time series.
\newblock {\em Chaos, Solitons $\&$ Fractals}, 104(Supplement C):323 -- 326,
  2017.

\bibitem{PRE}
L~G S ; Linhares C.A. ; da MOTA L. A. C.~P. BARBOSA, L. M. C. R. ;~DUARTE.
\newblock Improving the global fitting method on nonlinear time series
  analysis.
\newblock {\em Physical Review. E, Statistical, Nonlinear, and Soft Matter
  Physics (Print)}, 74:026702, 2006.

\bibitem{carli}
H.~Carli, L.G.S. Duarte, and L.A.C.P. da~Mota.
\newblock A maple package for improved global mapping forecast.
\newblock {\em Computer Physics Communications}, 185(3):1115 -- 1129, 2014.

\bibitem{bib:Boyce}
R.C. Boyce, W.E.;~Diprima.
\newblock {\em Equaç\~oes Diferenciais Elementares e Problemas de Valores de
  Contorno.}
\newblock LTC, Rio de Janeiro, 10 edition, 2015.

\bibitem{letellier2010influences}
Messager~V. Letellier, C.
\newblock Influences on otto e. r{\"o}ssler's earliest paper on chaos.

\bibitem{tigan2008analysis}
Gheorghe Tigan and Dumitru Opri{\c{s}}.
\newblock Analysis of a 3d chaotic system.
\newblock {\em Chaos, Solitons \& Fractals}, 36(5):1315--1319, 2008.

\bibitem{henon1964applicability}
Michel H{\'e}non and Carl Heiles.
\newblock The applicability of the third integral of motion: some numerical
  experiments.
\newblock {\em The Astronomical Journal}, 69:73, 1964.

\bibitem{barrow2003test}
John~D Barrow and Janna Levin.
\newblock A test of a test for chaos.
\newblock {\em arXiv preprint nlin/0303070}, 2003.

\bibitem{letellier2005relation}
Christophe Letellier, Luis~A Aguirre, and Jean Maquet.
\newblock Relation between observability and differential embeddings for
  nonlinear dynamics.
\newblock {\em Physical Review E}, 71(6):066213, 2005.

\bibitem{aguirre2008observability}
Luis~A Aguirre, Saulo~B Bastos, Marcela~A Alves, and Christophe Letellier.
\newblock Observability of nonlinear dynamics: Normalized results and a
  time-series approach.
\newblock {\em Chaos: An Interdisciplinary Journal of Nonlinear Science},
  18(1):013123, 2008.

\end{thebibliography}

\end{document}